\documentclass{article}
\usepackage{graphicx} % Required for inserting images
\usepackage{graphicx, wrapfig} % Required for inserting images
\usepackage{booktabs}
\usepackage{svg}
\usepackage{subfigure}
\usepackage{makecell}
\usepackage{threeparttable}
\usepackage{times}  
\usepackage[pdfpagelabels=false]{hyperref}
\usepackage{titlesec}
\usepackage{array, multirow}
\hypersetup{pdfstartview=FitH, pdfpagelayout=SinglePage}
\usepackage{colortbl}
\usepackage{booktabs}
\usepackage[normalem]{ulem}
\usepackage{moresize}
\usepackage{amsmath, amssymb, amsfonts}
\usepackage{algorithm}
\usepackage{algpseudocode}

\title{DESiRED - Dynamic, Enhanced, and Smart iRED: A P4-AQM with Deep Reinforcement Learning and In-band Network Telemetry}
\author{Leandro C. de Almeida, Washington Rodrigo Dias da Silva, \\ Thiago C. Tavares, Rafael Pasquini, Chrysa Papagianni and Fábio L. Verdi}

\begin{document}

\maketitle

\begin{abstract}
Active Queue Management (AQM) is a mechanism employed to alleviate transient congestion in network device buffers, such as routers and switches. Traditional AQM algorithms use fixed thresholds, like target delay or queue occupancy, to compute random packet drop probabilities. A very small target delay can increase packet losses and reduce link utilization, while a large target delay may increase queueing delays while lowering drop probability. Due to dynamic network traffic characteristics, where traffic fluctuations can lead to significant queue variations, maintaining a fixed threshold AQM may not suit all applications. Consequently, we explore the question: \textit{What is the ideal threshold (target delay) for AQMs?} In this work, we introduce DESiRED (Dynamic, Enhanced, and Smart iRED), a P4-based AQM that leverages precise network feedback from In-band Network Telemetry (INT) to feed a Deep Reinforcement Learning (DRL) model. This model dynamically adjusts the target delay based on rewards that maximize application Quality of Service (QoS). We evaluate DESiRED in a realistic P4-based test environment running an MPEG-DASH service. Our findings demonstrate up to a 90x reduction in video stall and a 42x increase in high-resolution video playback quality when the target delay is adjusted dynamically by DESiRED.
\end{abstract}

\section{Introduction}
\label{sec:introduction}

In the modern domain of computer networks, the necessity to meet progressively rigorous service requirements, including ultra-reliable low-latency communications and high bandwidth, has resulted in an unparalleled upsurge in network traffic, amplifying the intricacies associated with traffic management. Subsequently, approaches aimed at assisting congestion control mechanisms, such as Active Queue Management (AQM), are consistently embraced.

In scenarios where incoming packet rates exceed a network device's processing capacity, a transient queuing of packets occurs in the appropriate output queue, often causing transmission delays. To mitigate this bottleneck, an effective strategy involves notification congestion status to the packet sender, allowing the congestion control algorithm to reduce transmission rates. The primary methods for conveying congestion conditions to senders include packet marking using Explicit Congestion Notification (ECN) bits and selective packet dropping. These approaches are the predominant means of communicating congestion information in network environments.

Traditionally, AQM mechanisms have been primarily focused on draining packets directly from queues, with the overarching objective of mitigating transient congestion occurrences and reducing the queuing delay. Prominent examples of these traditional AQM algorithms include Random Early Detection (RED) \cite{RED:93}, Blue \cite{Blue:2002}, CoDel \cite{CODEL:2012}, CAKE \cite{CAKE:2018}, and PIE \cite{ietf-aqm-pie-03}. More recently, owing to the inherent flexibility of the programmable data plane (PDP), the prevailing state-of-the-art AQM solutions designed to operate within PDP hardware environments and made publicly accessible comprise iRED \cite{iRED:2022}, P4-CoDel \cite{kundel2021p4codel}, and the (dual) PI2 \cite{DualPI2}. These AQM implementations exemplify the synergy between novel programmable data plane capabilities and the evolving demands of congestion control within modern network infrastructures.

\begin{figure}[ht]
    \centering
    \includegraphics[width=.7\columnwidth]{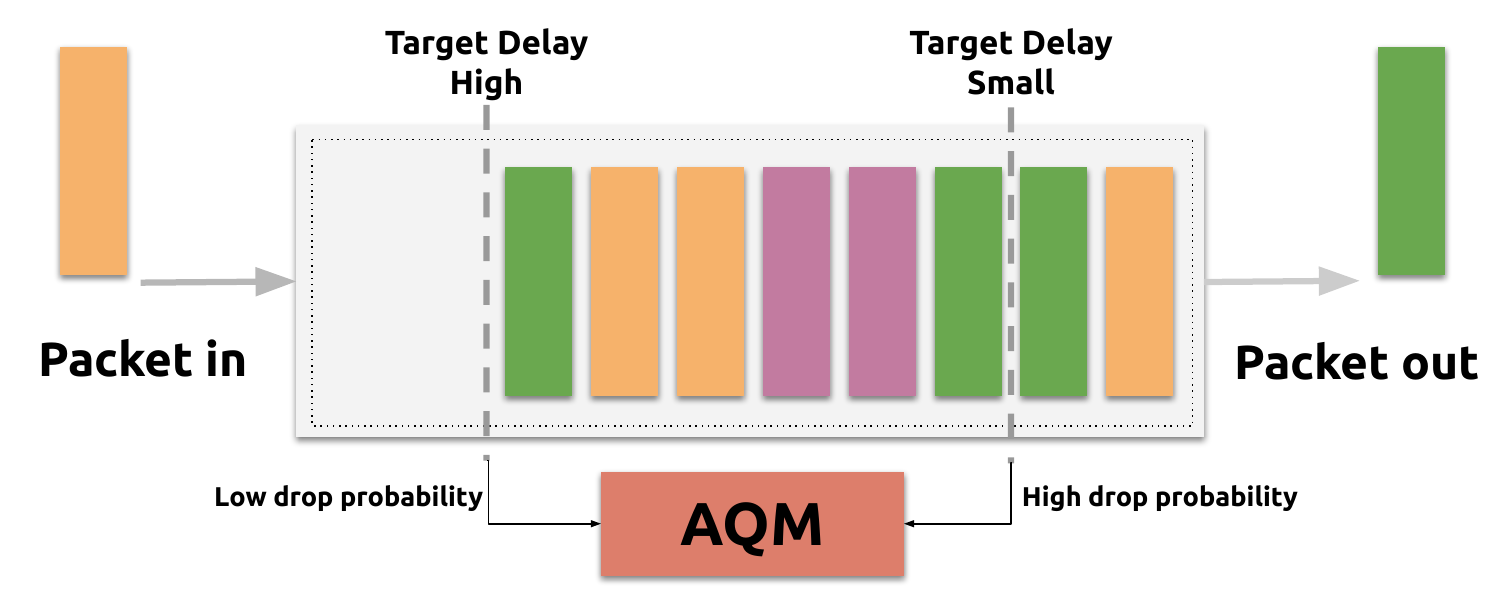}
    \caption{The trade-off: If the Target Delay is small, can increase packet losses and decrease link utilization. If is high, increase queueing delays and decrease packet drops.}
    \label{fig:fixed-tdprob}
\end{figure}

An integral aspect of AQM algorithms pertains to the selection of an appropriate threshold value, often determined based on considerations of either queue delay (referred to as the target delay) or queue depth. Opting for an excessively small threshold value can lead to an increased occurrence of packet losses, resulting in a higher drop probability while reducing overall link utilization. Conversely, employing a high threshold value can lead to extended queuing delays but a lower likelihood of packet drops, characterized by a reduced drop probability. Additionally, the dynamic nature of network traffic necessitates the avoidance of static threshold values for specific applications. In this context, we explore this issue as the \textbf{fixed target delay problem}, as illustrated in Fig. \ref{fig:fixed-tdprob}, delving into the intricate dynamics of threshold determination in AQM algorithms

%In Fig \ref{fig:fixed-tdprob}, the AQM algorithm continuously monitors whether packets are exceeding thresholds predefined by the network administrator, either a small threshold (Target Delay Small) or a high threshold (Target Delay High). In the case of the former, stringent control is promoted over the number of packets within the queue, resulting in a big incidence of packet drops. The primary advantage of this approach lies in its adeptness at managing queuing delays. However, this advantage is compensated by the substantial number of dropped packets, necessitating their retransmission and thereby imposing a penalty on the application. On the other hand, the latter approach, characterized by a high target delay, yields fewer packet exceedances of the threshold. Consequently, a diminished occurrence of packet drops ensues, but correspondingly, the queuing delay tends to escalate commensurate with the intensity of network traffic.

At the core of this matter lies a fundamental trade-off, giving rise to a pivotal question: \textit{What is the ideal target delay for AQM?} Estimating this value presents a challenging task. However, recent advancements in the field of artificial intelligence as applied to computer networks \cite{ML4Net:2018} introduce a potential avenue, leveraging the capabilities of Deep Reinforcement Learning (DRL) as a powerful tool to enhance decision-making in addressing this challenge.

Although DRL models are known for their appetite for data, the provision of real-time data at the requisite granularity has posed an obstacle within the realm of computer networks. However, recent advances in the domain of PDP, in tandem with the integration of In-band Network Telemetry (INT) \cite{INT:2021}, have conducted a paradigm shift. These advancements have presented us with the capability to attain granular visibility, discernible on a per-packet basis, effectively altering the scenario of the challenges associated with data availability in the context of DRL applications within computer networks.

The hypothesis of this study posits that INT measurements can serve as valuable input features for a DRL model. This DRL model is intended to dynamically adjust the target delay, departing from our prior work with a fixed target delay in iRED \cite{iRED:2022}. The overarching goal is to utilize this DRL model for real-time optimization of QoS, thereby introducing a novel approach aimed at enhancing network performance and adaptability.

iRED represents a pioneering P4-based algorithm that introduced the concept of disaggregated AQM in PDP hardware. Disaggregated AQM involves the segmentation of AQM operations into distinct blocks, specifically Ingress and Egress, within the PDP architecture. Addittionally, iRED achieves full compliance with the L4S framework (Low Latency, Low Loss, and Scalable Throughput)  \cite{L4S:2023}. It accomplishes this by categorizing traffic as either Classic (subject to dropping) or Scalable \footnote{TCP Prague in L4S framework.} (marked with the ECN bit), thus ensuring fairness among various flows through a combined packet dropping and marking mechanism.

Through the integration of INT, DRL, and the iRED framework, we introduce the innovative paradigm of DESiRED (Dynamic, Enhanced, and Smart iRED). To our knowledge, DESiRED serves as the leading implementation of a dynamic AQM system based on P4 architecture. This advancement combines the cutting-edge capabilities of fine-grained network measurements enabled by INT with the cognitive capabilities provided by the Deep Q-Network (DQN), thereby representing an integrated embodiment of state-of-the-art progress in the field of AQM.

\begin{figure}[ht]
    \centering
    \includegraphics[width=.8\columnwidth]{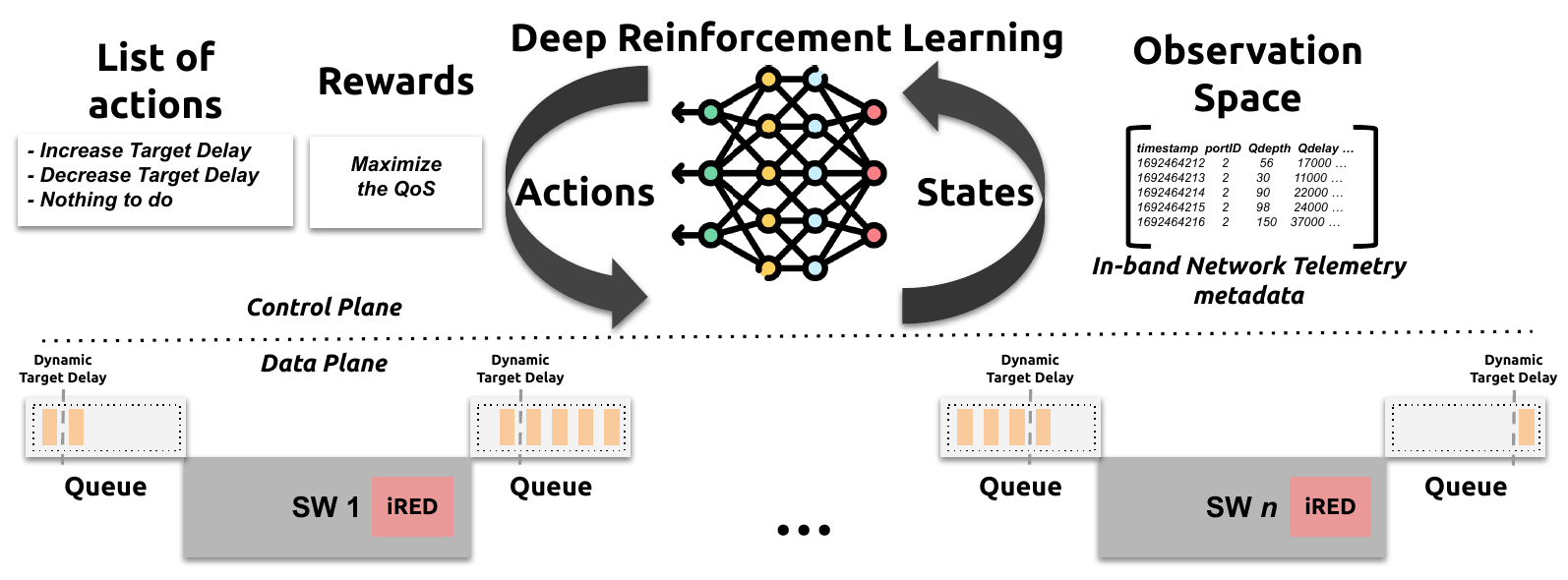}
    \caption{The Smart Control Closed Loop with DESiRED: The fine-grained INT measurements provide wide visibility into the state of the network in each observation space. The DRL mechanism guides the agent to dynamically adjust the target delay to maximize the QoS.}
    \label{fig:DQN-RL}
\end{figure}

We undertake a comprehensive evaluation of DESiRED within a realistic testbed environment, focusing on the delivery of an MPEG-DASH (Dynamic Adaptive Streaming over HTTP) service \cite{DASH:2014}. Our experiments involve the provision of diverse video catalogs to video clients traversing a programmable network. Fine-grained INT measurements, collected at line rate in the data plane, are utilized to inform the DRL mechanism in the control plane. The DRL mechanism guides the agent's actions, dynamically adjusting the target delay to optimize the QoS for the DASH service. This forms a Smart Control Closed Loop, as depicted in Fig. \ref{fig:DQN-RL}

Our empirical findings elucidate that DESiRED wields an impact, with the potential to alleviate video stall occurrences by a factor of up to 90x. Moreover, the enhancement in the QoS within the MPEG-DASH framework is evident, as measured by an augmentation of up to 42x in terms of Frames per Second (FPS), underscoring the considerable efficacy of DESiRED in elevating the video streaming experience.
In summary, the main contributions of this work are:

\begin{enumerate}
    \item We design and implement DESiRED, which is a smart Closed Control Loop that unifies the state of the art in network telemetry (INT), Deep Reinforcement Learning (DQN), and congestion control in-network (AQM).
    \item We conduct an evaluation of the DESiRED algorithm within the context of a DASH service. This entails the practical implementation of DESiRED within a real-setup DASH environment, followed by a systematic evaluation of its performance and effectiveness.
    \item We have created and made publicly available datasets used throughout our experiments that encompass network and application data, collectively characterizing the complexities of an adaptive video service. 
\end{enumerate}

The remainder of the paper is organized as follows. In Section 2 we describe INT and DRL fundamental concepts. Additionally, we detail DESiRED, describing the main components implemented in the P4 language (data plane) and the DRL integration (control plane) in Section 3. In Section 4, the experiments and evaluation are presented, including a brief view of the testbed and the workloads used. Results and discussion are detailed in Section 5. Some Lessons learned are given in Section 6. Finally, the conclusions are depicted in Section 7.

\section{Background}

In this section, we expound upon the foundational principles that underpin the functionality of DESiRED. Sub-section \ref{sec:INT} provides a concise elucidation of the programmable data plane and In-band Network Telemetry. Furthermore, Section \ref{sec:DRL} delves into the principal facets of Deep Reinforcement Learning.

\subsection{In-band Network Telemetry} \label{sec:INT}

Recent progress in programmable hardware and the utilization of the P4 language \cite{P4paper:2014} have enabled network devices to autonomously report the network's state, eliminating the need for direct control plane intervention \cite{McKeown:2019}. In this scenario, packets incorporate telemetry instructions within their header fields, facilitating the fine-grained collection and recording of network data. 

\begin{figure}[ht]
    \centering
    \includegraphics[width=.7\columnwidth]{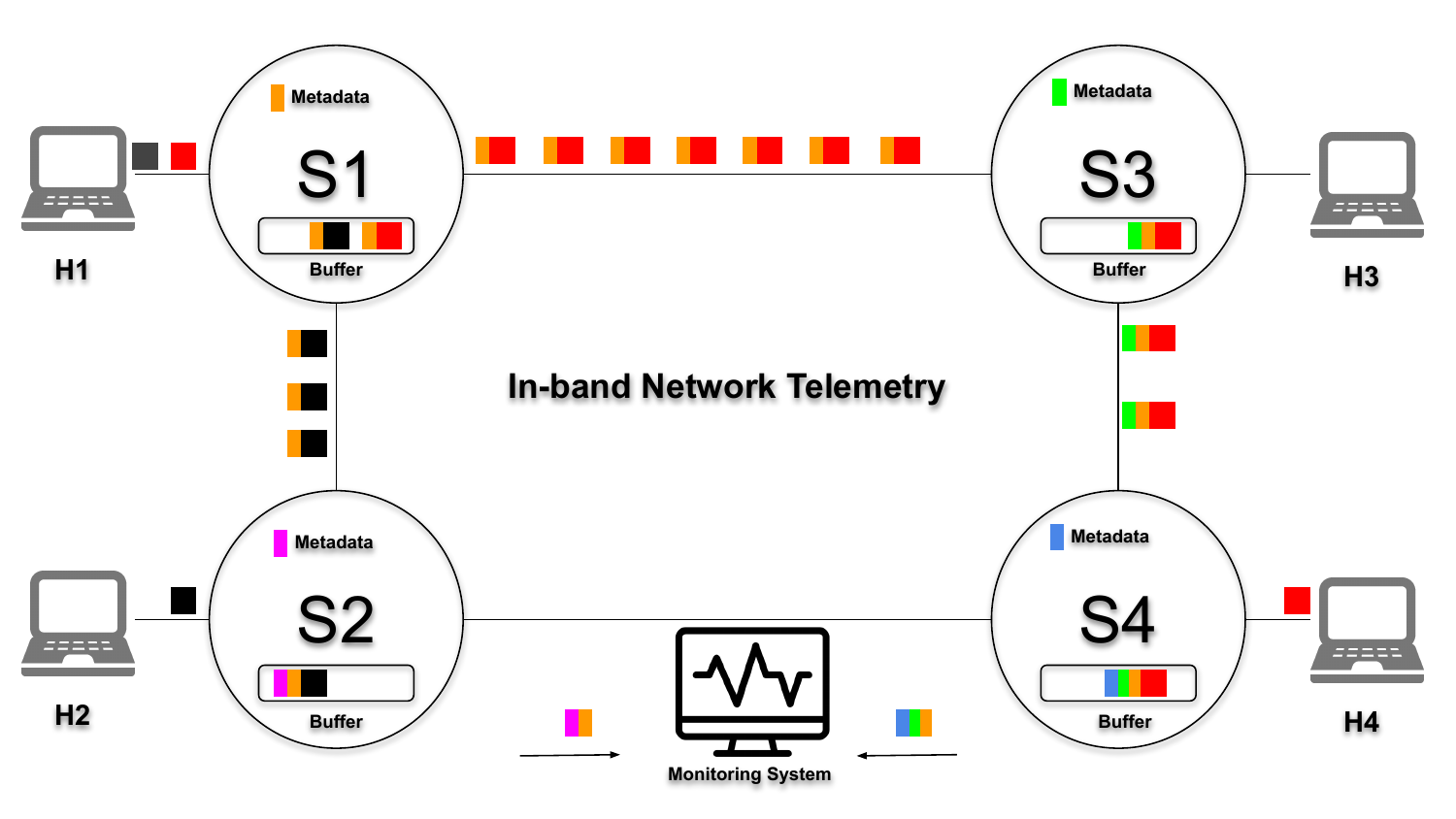}
    \caption{In-band Network Telemetry operation. INT metadata is appended on the packets in each hop. In the specific collection point, the monitoring system receives INT metadata.}
    \label{fig:INT}
\end{figure}

The telemetry instructions are defined in the INT data plane specification \cite{INT:2021}. Figure \ref{fig:INT} illustrates the operation of INT within an arbitrary network. The network comprises four end systems, namely \textit{H1, H2, H3}, and \textit{H4}, along with four nodes equipped with P4 and INT support, denoted as \textit{S1, S2, S3}, and \textit{S4}. Each network node possesses a set of metadata, represented by orange (S1), magenta (S2), green (S3), and blue (S4) rectangles. This metadata contains information specific to each node, such as Node ID, Ingress Port, Egress Spec, Egress Port, Ingress Global Timestamp, Egress Global Timestamp, Enqueue Timestamp, Enqueue Queue Depth, Dequeue Timedelta, and Dequeue Queue Depth, as specified in the V1Model architecture.

%, which delineates three distinct modes of operation: INT-XD (\textit{eXport Data}), INT-MX (\textit{eMbed Instructions}), and INT-MD (\textit{eMbed Data}).

%In the INT-XD mode, each device exports metadata by utilizing its INT instructions, which are configured in flow tables directly within the data plane, transmitting the information to the monitoring system. In this mode, no alterations are made to the application traffic.

%In the INT-MX mode, the source node generates INT instructions within the packet header. At each transit node, these instructions are read, and the corresponding metadata is directly transmitted to the monitoring system. In this mode, minor modifications are introduced to the application traffic, as only telemetry instructions are inserted into the packet headers.

%In the INT-MD mode, both INT instructions and metadata are inserted into packets at each network hop. This represents the default operating mode as defined in the INT specification. For clarity, we will refer to the INT-MD mode as simply ``INT" in this paper.

In Figure \ref{fig:INT}, there are two distinct flows depicted: one represented by red packets and the other by black packets. The red flow is required to adhere to the prescribed network path \textit{f1={H1, S1, S3, S4, H4}}, while the black flow must traverse the designated path \textit{f2={H1, S1, S2, H2}}.

At each network hop along these paths, the data plane of the network devices employs telemetry instructions to facilitate the collection and inclusion of metadata within the packets as they traverse each node. This process is iteratively performed throughout the journey, starting from the first node after the source and concluding at the last node before reaching the destination. Upon reaching the destination node, the metadata is extracted from the packet and subsequently relayed to the monitoring system. The original packet is then directed to its final destination.

In addition to the modes delineated in the INT specification, alternative approaches exist for collecting metadata within programmable networks. One such approach involves the utilization of an ``exclusive telemetry flow" to monitor the network's state, which, in this work, is referred to as ``Out-of-band Network Telemetry" (ONT).

In the ONT scenario, dedicated probe packets are employed to gather metadata, eliminating the need for any modifications to the data packets associated with the services operating within the network. The primary advantage of this approach lies in its ability to maintain the integrity of application traffic, as it traverses the programmable network without undergoing alterations, thereby mitigating issues related to packet growth, such as fragmentation.

Conversely, the use of an exclusive telemetry flow introduces additional overhead to the overall network traffic. This is due to the necessity of having a dedicated monitoring flow ONT for each service running within the network.

One of the primary advantages of employing telemetry lies in the exceptional level of granularity it offers. Every individual packet traversing the network carries pertinent information directly to the monitoring system at the line rate. This level of granularity aligns with the perspective presented in \cite{Orch-INT-ML}, wherein it is recognized that a substantial volume of data can prove immensely valuable for Deep Reinforcement Learning (DRL) algorithms, which have a voracious appetite for information.

\subsection{Deep Reinforcement Learning} \label{sec:DRL}

% 1st Version
%Reinforcement Learning (RL) constitutes an Artificial Intelligence (AI) learning paradigm centered on actions and rewards. Unlike the conventional supervised and unsupervised learning approaches, where models learn from predefined dataset features, an RL learner, also known as an agent, engages in interaction with an environment and receives rewards or penalties based on the actions it takes.

Reinforcement Learning (RL) is an Artificial Intelligence (AI) learning paradigm centered on actions and rewards. Unlike the conventional supervised and unsupervised learning approaches, where models learn from predefined dataset features, an RL learner, also known as an agent, interacts with an environment and receives rewards or penalties based on the actions it takes.

\begin{figure}[htb!]
    \centering
    \includegraphics[width=.8\columnwidth]{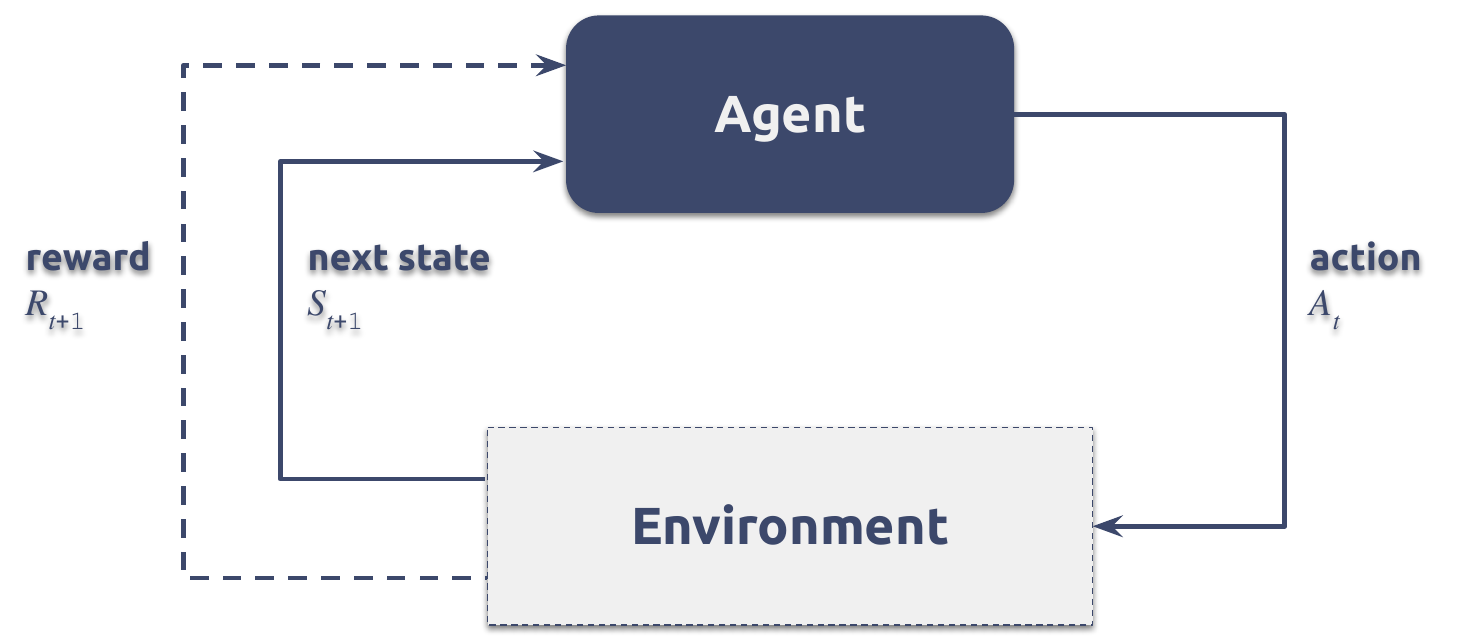}
    \caption{Interaction between the agent and environment in a Markov Decision Process. Adapted from \cite{sutton2018reinforcement}.}
    \label{fig:rl_agent-env_interaction}
\end{figure}

The model depicted in Figure \ref{fig:rl_agent-env_interaction} illustrates the formalization of a sequential decision-making strategy known as a Markov Decision Process (MDP). In this framework, the agent continually interacts with the environment by executing actions ($A$) at specific time steps ($t$) and observing new states ($S_{t+1}$) resulting from these actions. After each interaction, a reward value ($R_{t+1}$) is generated to assess the correctness of the action, with the aim of maximizing cumulative rewards throughout the agent's training process \cite{boutaba2018comprehensive, sutton2018reinforcement}.

% [ref. 1]: A comprehensive survey on machine learning for networking: evolution, applications and research opportunities
% https://jisajournal.springeropen.com/articles/10.1186/s13174-018-0087-2#Sec2 

% [ref. 2]: Reinforcement Learning: An Introduction
% https://www.andrew.cmu.edu/course/10-703/textbook/BartoSutton.pdf

In this context, the agent learns to maximize its cumulative rewards by determining a policy\footnote{A policy defines the agent's strategy for associating actions with states. Such strategies can be stochastic, specifying probabilities for each action that could be taken when a particular state is observed \cite{sutton2018reinforcement}.} that optimizes an action-value function, denoted as $Q$. This function estimates the quality of actions taken by the agent in specific states.

Formally, an optimal action-value function, denoted as $q_*$, can be defined using the Bellman optimality equation \cite{sutton2018reinforcement}:

\begin{equation}\label{eq:bellman_optimal_q-value}
    q_*(s,a) = \mathbb{E}[R_{t+1} + \gamma\, \underset{a'}{\max}\, q_*(S_{t+1},a') \ | \ S_t = s, A_t = a]
\end{equation}

Intuitively, the Bellman optimality equation suggests that the optimal Q-value for any state-action pair ($q_*(s,a)$) is expected to be the immediate reward obtained after taking an action ($a$) in a given state ($s$) at time step ($t$), augmented by the maximum expected return achievable by adhering to an optimal policy for subsequent state-action pairs, which are discounted\footnote{This discounting approach allows the agent to prioritize actions that maximize the cumulative rewards it receives in the future. The discount rate, denoted as $\gamma$, determines the present value of future rewards. For instance, a reward received $k$ time steps in the future is worth only $\gamma^{k-1}$ times what it would be worth if it were received immediately \cite{sutton2018reinforcement}.} by $\gamma$ \cite{sutton2018reinforcement}.

Hence, the resolution of the Bellman optimality equation provides a pathway to ascertain an optimal policy, offering a potential solution to a RL problem. Nevertheless, it is imperative to acknowledge that, in practice, this solution is seldom feasible. It resembles an exhaustive search that requires consideration of all possible scenarios, involving the computation of occurrence probabilities and expected reward returns. Additionally, it relies on three assumptions that are often challenged when implementing solutions for real-world problems:

a) The accurate knowledge of environmental dynamics.

b) Sufficient computational resources to complete the computational requirements of the solution.

c) The adherence to the Markov property.

In light of these challenges, the only pragmatic approach to tackle the Bellman optimality equation is to seek a policy approximation derived from actual experiences, where transitions involving state $s$, action $a$, and reward $r$ are considered, as opposed to relying solely on the expected outcomes \cite{sutton2018reinforcement}.

When dealing with scenarios characterized by a well-defined set of finite states, it becomes feasible to model an approximation of the Bellman optimality equation using tabular data structures. Each entry within these structures corresponds to a state-action pair. In this context, the Q-Learning algorithm, introduced by Watkins \cite{watkins1992q}, marked a significant milestone in the early stages of the RL paradigm.

Q-Learning is noteworthy for its direct approximation of the Bellman optimality equation, irrespective of the policy in use. It simplified the analysis of agent algorithms and facilitated early convergence proofs \cite{sutton2018reinforcement}. The Watkins Q-Learning algorithm is formally defined as follows:

% [ref. 3]: Q-Learning
% https://link.springer.com/content/pdf/10.1007/BF00992698.pdf

\begin{equation}\label{eq:q-learning}
    Q(S_t,A_t) = (1 - \alpha)Q(S_t,A_t) \ + \ \alpha[R_{t+1} \ + \ \gamma \ \underset{a}{\max} \ Q(S_{t+1},a)]
\end{equation}

\noindent{where the approximated optimal Q-value is calculated by blending the current Q-value, denoted as $Q(S_t, A_t)$, with the target temporal difference. The target temporal difference represents the reward $R_{t+1}$ obtained when transitioning to the subsequent state $S_{t+1}$ after taking action $a$. This value is then weighted by the discount factor $\gamma$ and modulated by a learning rate $\alpha$ ($0 \leq \alpha < 1$) \cite{watkins1992q, sutton2018reinforcement}.}

However, it is essential to acknowledge that Q-learning operates under the assumption of a tabular representation for state-action pairs and approximates the optimal Q-value in a linear fashion. In practice, real-world applications often exhibit complexity, characterized by non-linear relationships and encompassing high-dimensional state spaces. Such complexities render the storage of comprehensive tables unfeasible \cite{sutton2018reinforcement}.

\textit{Networking management serves as a compelling example of such scenarios, where modern Tofino switches can process INT packets at a nanosecond timescale.}

To address these limitations, Mnih et al. leveraged the Q-Learning algorithm by integrating it with a Deep Neural Network (DNN) to approximate the optimal Q-value,  a methodology known as Deep Q-Network (DQN). In their seminal work \cite{mnih2015human}, the authors showcased the effectiveness of this approach by training and evaluating the DQN on an Atari 2600 emulator. Impressively, the DQN-based agents achieved performance levels surpassing those of human players in 49 distinct games, relying solely on pixel inputs and game scores for guidance.

Of note, the authors maintained a consistent algorithm, DNN architecture, and hyperparameters across all games, eliminating the need for game-specific feature engineering. Thus, DQN not only outperformed agents employing linear function approximation but also demonstrated the capacity to attain or exceed human-competitive skills across diverse gaming environments. This pioneering work exemplified the synergy between RL and contemporary Deep Learning (DL) techniques, signifying a significant advancement in the state of artificial intelligence. It underscored the potential of RL when combined with modern DL methods, yielding remarkable outcomes \cite{mnih2015human, sutton2018reinforcement}.

% [ref. 4]: Human-level control through deep reinforcement learning
% https://storage.googleapis.com/deepmind-media/dqn/DQNNaturePaper.pdf

In line with this, we present an RL-based approach designed to dynamically fine-tune the iRED target delay to an optimal value during video streaming, named DESiRED. This process is facilitated by an agent built on the foundation of DQN. In the subsequent Subsection, we will delve into the constituent elements that constitute this innovative approach.

\subsubsection{Deep Q-Network workflow}\label{subsubsec:dqn}

%The DQN architecture proposed by [ref. 4] comprises a Deep Convolutional Neural Network (CNN) that receives the emulated game frames as input, and outputs the predicted Q-values for each possible action that could be taken for the given input state. In order to approach the solution proposed in this paper, we modeled the DQN using the Multi-layer Perceptron (MLP) architecture due to the INT metrics tabular characteristic. A MLP network encompass an input layer comprising units for each feature, $n$ hidden layers with $m$ Rectified Linear Units (ReLU) and an output layer containing units for each action the agent could take (see subsection \ref{subsec:ired-dqn} for more details).

The DQN architecture, as proposed by Mnih et al. \cite{mnih2015human}, consists of a Deep Convolutional Neural Network (CNN) designed to receive emulated game frames as input and subsequently generate predicted Q-values for each potential action within the given input state. To facilitate such predictions, Mnih et al. introduced two critical modifications to the conventional Q-Learning algorithm. These alterations were essential to mitigate instabilities inherent in using Deep Neural Networks (DNNs) for Q-value approximation \cite{sutton2018reinforcement}.

The first modification entails the incorporation of a biologically inspired mechanism referred to as 'experience replay.' In this approach, the agent's experiences are stored as tuples containing the current state ($S_t$), the action taken ($A_t$), the reward received ($R_{t+1}$), and the subsequent state ($S_{t+1}$). Periodically, after reaching a predefined replay memory limit, a mini-batch of these experiences is uniformly sampled for training the DNN \cite{mnih2015human, sutton2018reinforcement}.

\begin{figure}[htb!]
    \centering
    \includegraphics[width=.8\columnwidth]{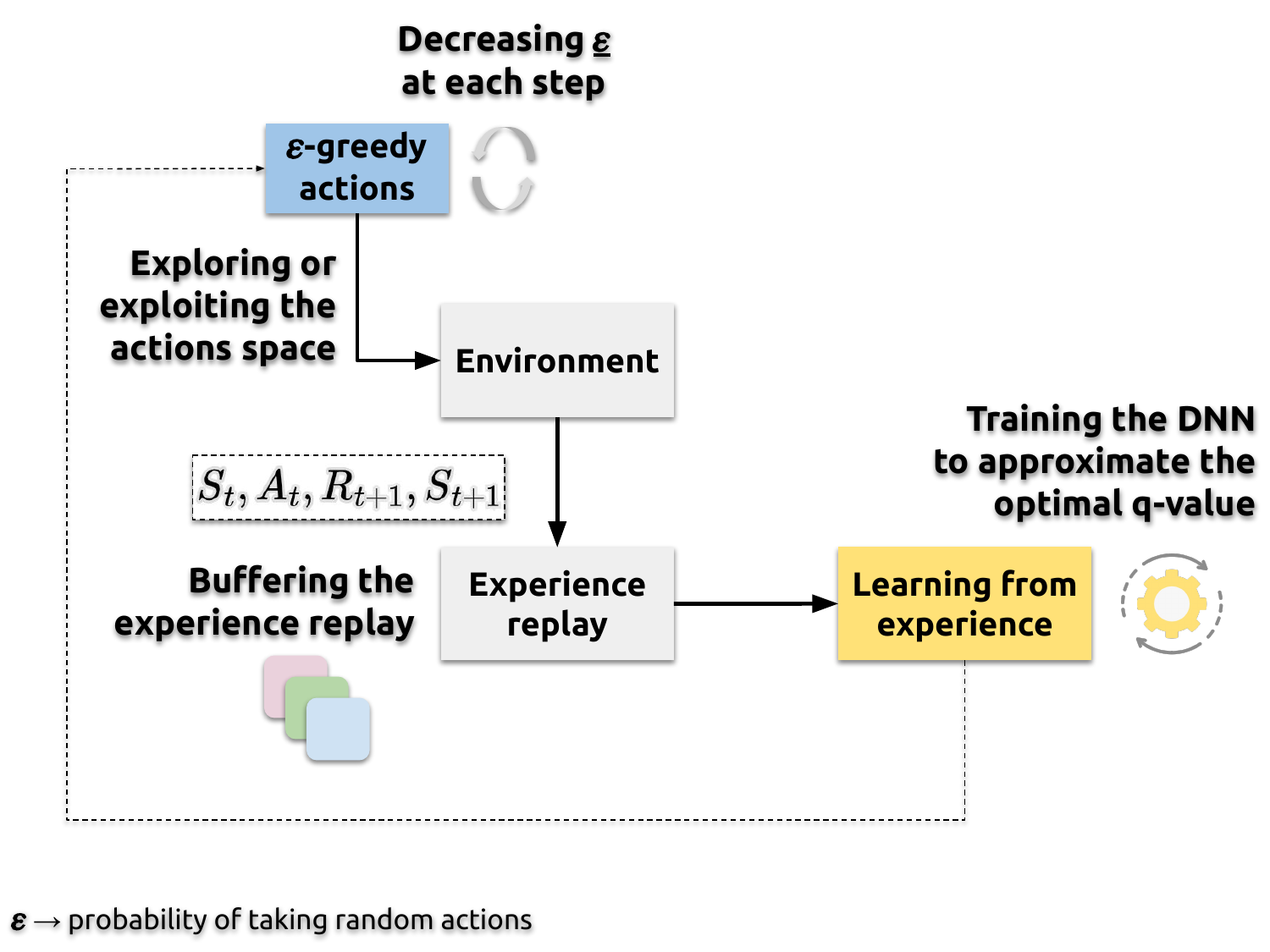}
    \caption{DQN high-level workflow \cite{mnih2015human, sutton2018reinforcement}.}
    \label{fig:rl_dqn-env_interaction}
\end{figure}

This approach plays a pivotal role in mitigating the emergence of correlations within the observed state space. By decoupling the dependence on successive experiences, it effectively reduces the variance in the parameters of the DNN. Fig. \ref{fig:rl_dqn-env_interaction} illustrates the interaction between a DQN agent and an environment, taking into account the experience replay mechanism. Within this context, the agent selects actions following an $\epsilon$-greedy rule.

Specifically, when employing this rule, the agent chooses between two strategies: ``exploitation" and ``exploration". A ``greedy action" involves selecting an action from the action space based on the maximum estimated Q-value. Conversely, a ``non-greedy action" entails the random selection of an action. Exploitation, represented by the selection of a greedy action, aims to exploit the current knowledge to maximize immediate rewards. In contrast, exploration, represented by non-greedy actions, focuses on traversing the action space to maximize cumulative rewards in the long run \cite{sutton2018reinforcement}.

In RL, achieving a balanced trade-off between exploration and exploitation is paramount. However, it's important to acknowledge that, at a single time step, it's not possible for an agent to simultaneously exploit and explore actions. To reconcile these opposing strategies, a solution is to allow the agent to primarily act greedily, favoring exploitation, while intermittently choosing an action from the action space at random, independent of the estimated Q-values. This random selection is determined by an exponentially decreasing probability parameter $\epsilon$. Consequently, as the time steps progress, the probability of selecting an optimal action gradually converges to a value greater than $1-\epsilon$, approaching near certainty in favor of exploitation as the agent refines its strategy over time \cite{sutton2018reinforcement}.

%If you maintain estimates of the action values, then at any time step there is at least one action whose estimated value is greatest. We call these the greedy actions. When you select one of these actions, we say that you are exploiting your current knowledge of the values of the actions.  If instead you select one of the nongreedy actions, then we say you are exploring, because this enables you to improve your estimate of the nongreedy action’s value. Exploitation is the right thing to do to maximize the expected reward on the one step, but exploration may produce the greater total reward in the long run.

%Greedy action selection always exploits current knowledge to maximize immediate reward; it spends no time at all sampling apparently inferior actions to see if they might really be better. A simple alternative is to behave greedily most of the time, but every once in a while, say with small probability e, instead select randomly from among all the actions with equal probability, independently of the action-value estimates. An advantage of these methods is that, in the limit as the number of steps increases, every action will be sampled an infinite number of times, thus ensuring that all the Qt(a) converge to q*(a). This of course implies that the probability of selecting the optimal action converges to greater than 1 - e, that is, to near certainty.

%Reinforcement learning requires a balance between exploration and exploitation

A second significant contribution introduced by Mnih et al. \cite{mnih2015human}, relative to classical Q-Learning, pertains to the learning stage of the DQN. In this stage, a separate network, referred to as the 'target network,' is employed to estimate target values for the Q-network, often referred to as the 'online network.' This modification enhances the algorithm's stability compared to using a single online network. The rationale behind this improvement lies in the fact that updating the parameters of the online network for the current state-action pair can inadvertently influence the Q-values of the next state, potentially leading to oscillations or even policy divergence.

To address this challenge, the online network's parameters are periodically cloned to the target network at intervals of every $C$ time steps. Consequently, the target network's predictions serve as target values for the online network during the subsequent $C$ time steps. This introduces a delay in updating the Q-values between the current and next states, effectively reducing the likelihood of policy oscillations or divergence \cite{mnih2015human}.

\begin{figure}[htb!]
    \centering
    \includegraphics[width=.8\columnwidth]{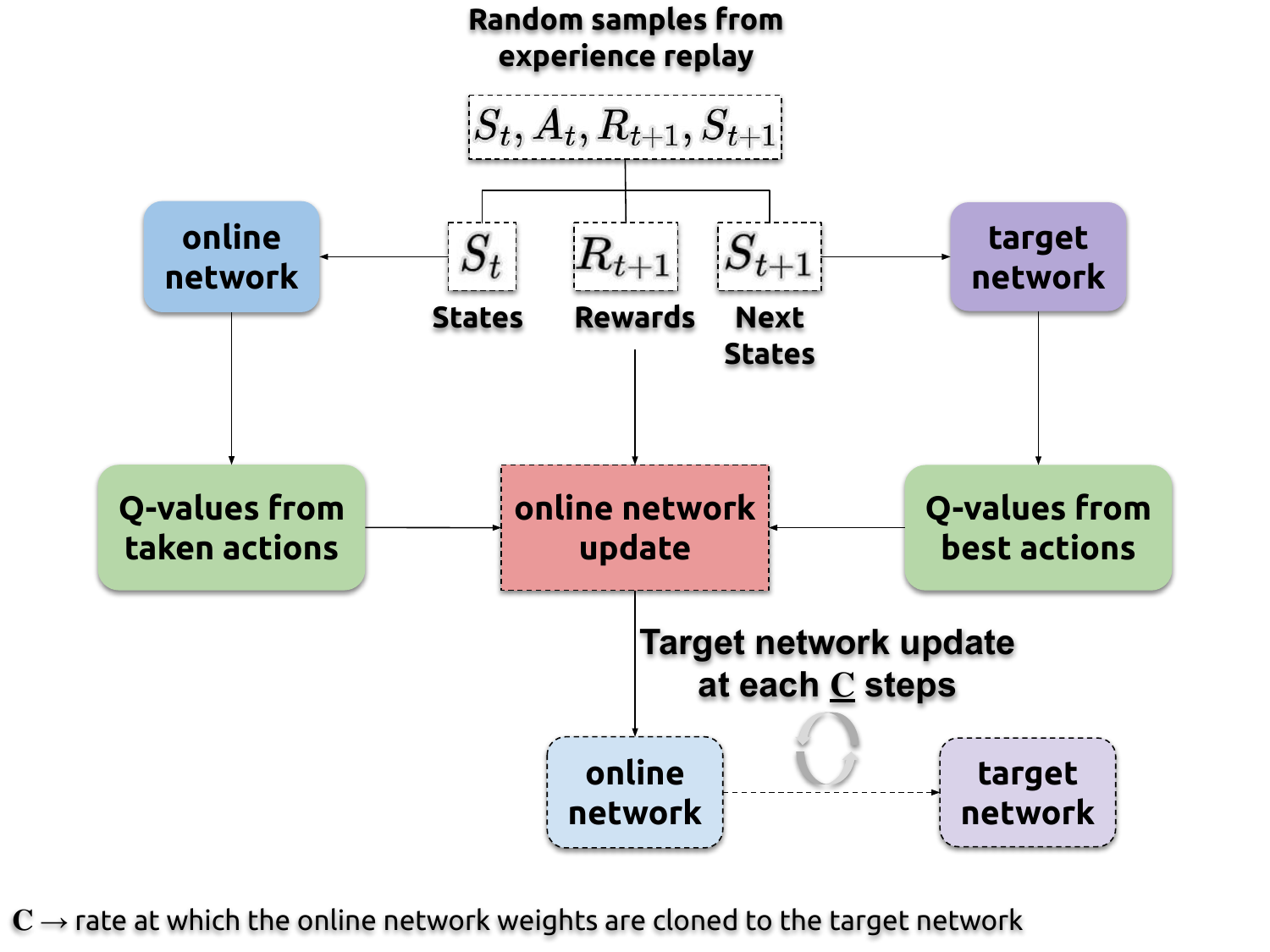}
    \caption{DQN learning stage workflow \cite{mnih2015human}.}
    \label{fig:rl_dqn-double_q-learning}
\end{figure}

Figure \ref{fig:rl_dqn-double_q-learning} illustrates the DQN learning workflow, incorporating the approach described above. A concise introduction to the functionality of DQN is presented in Subsection \ref{sec:CPDQN}. Furthermore, for a detailed exposition on the implementation of DQN within the scope of this research, please refer to Subsection \ref{subsec:ired-dqn}.

%In the subsequent subsection (\ref{subsec:ired-dqn}), we will elaborate on the modifications made to the foundational DQN architecture and workflow to enable iRED to dynamically adjust the target delay.

%The second modification to online Q-learning aimed at further improving the stability of our method with neural networks is to use a separate network for generating the targets yj in the Q-learning update. More precisely, every C updates we clone the network Q to obtain a target network Q^ and use Q^ for generating the Q-learning targets yj for the following C updates to Q. This modification makes the algorithm more stable compared to standard online Q-learning, where an update that increases Q(st,at) often also increases Q(st+1,a) for all a and hence also increases the target yj, possibly leading to oscillations or divergence of the policy. Generating the targets using an older set of parameters adds a delay between the time an update to Q is made and the time the update affects the targets yj, making divergence or oscillations much more unlikely

\section{DESiRED - Dynamic, Enhanced, and Smart iRED.}

DESiRED, herein referred to as an advanced iteration of iRED, which was initially introduced in the work of \cite{iRED:2022}, constitutes a notable enhancement within the realm of intelligent network control systems. Specifically, it introduces a novel capability wherein the intelligent control plane harnesses the power of DRL to dynamically optimize and fine-tune the target delay parameters. In alignment with its predecessor, iRED, DESiRED remains faithful to the fundamental concept of disaggregated AQM. In this paradigm, AQM operations are compartmentalized into discrete functional blocks within the architecture.

The concept of disaggregation emerges from the imperative to expedite packet discarding processes. In the pursuit of resource efficiency, we contend that the optimal location for packet discarding is the Ingress block. However, a noteworthy challenge arises as the vital metadata pertaining to queue delay (or queue depth), which constitutes the primary information utilized as input for the AQM algorithm to determine packet discarding decisions, is captured by the Traffic Manager and traditionally accessible within the Egress block. Within this context, DESiRED leverages a congestion notification mechanism, designed to incur minimal overhead, in order to relay the imperative to execute packet discarding actions to the Ingress block.

\begin{figure}[ht]
    \centering
    \includegraphics[width=1\columnwidth]{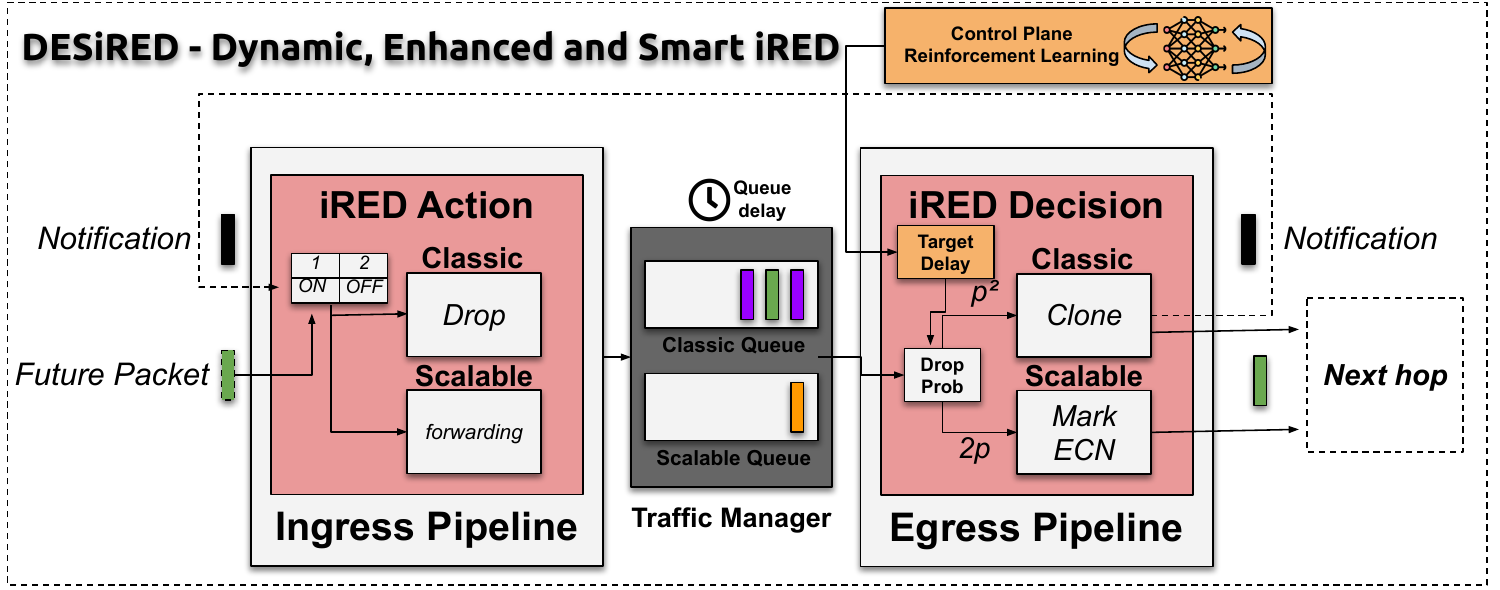}
    \caption{The Closed Control Loop overview with DESiRED. At the control plane side, DRL updates the target delay at the data plane.}
    \label{fig:iRED-RL}
\end{figure}

As illustrated in Figure \ref{fig:iRED-RL}, the decision-making process within DESiRED takes place at the Egress block, while the corresponding actions are subsequently executed at the Ingress block. The following Subsections will elucidate the functioning of DESiRED, with a distinct focus on data plane and control plane operations.

\subsection{Data plane operation (AQM)}

To provide a more comprehensive understanding, we will commence our description of DESiRED's operation from the data plane perspective, focusing initially on the Egress block. Our exploration will initiate with the drop or marking decision process, a critical component housed within the decision module. At the Egress, iRED computes the Exponentially Weighted Mean Average (EWMA) of the queue delay (or queue depth\footnote{The programmer can choose whether to use DESiRED's delay-based or depth-based approach.}) for each individual packet, entirely within the data plane. The inherent absence of division and floating-point operations poses challenges in calculating average values within the data plane. To surmount this limitation, as applied in \cite{bussegrawitz2022pforest}, we employ an approximation method following Eq. \ref{eq:ewma}:

\begin{equation}
    \centering
    S_t = \alpha \cdot Y_t + (1 - \alpha)\cdot S_{t-1}
    \label{eq:ewma}
\end{equation}

where $S_t$ is the updated average queue delay, $S_{t-1}$ is the previous average queue delay and $Y_t$ is the current queue delay. The constant $\alpha \in [0,1]$ determines how much the current value influences the average. We use $\alpha=0.5$, such multiplication can be replaced by bit shifts operations. The output of the EWMA will represent the average queue delay over time. When the observed value, representing the average queue delay, falls within the range (minimum and maximum thresholds) configured by the DRL mechanism, DESiRED proceeds to calculate the drop probability in accordance with the RED approach. Simultaneously, it employs a coupling mechanism to generate various levels of congestion signals, which may entail either packet drops or packet marking (ECN bit).

Once the DESiRED decision module (Egress) has detected that a packet must be dropped, DESiRED must notify the action module (Ingress) to perform this action. The first challenge in the PDP context is to achieve communication between the Ingress and Egress blocks with minimum overhead. Obviously, DESiRED will not drop the packet that generated the discard decision, but a future packet \cite{conquest:2019}. Discarding future packets is one of the main features differentiating DESiRED from other state-of-the-art AQMs. For the congestion notification to reach the Ingress block, DESiRED creates a congestion notification packet (clone packet) and sends it through an internal recirculation port to reach the Ingress block. 

The action module, situated in the Ingress block, maintains the congestion state table on a per-port/queue basis and activates the drop flag (ON) for the corresponding port/queue. The current packet is forwarded to the next hop without introducing any additional delay. Subsequently, future packets intended for the same output port/queue, where the drop flag is set to ON, will be dropped, and the drop flag will be reset to OFF. This mechanism, facilitated by DESiRED, ensures that the Ingress pipeline can proactively mitigate imminent queue congestion.

\subsection{Control plane operation (DRL)} \label{sec:CPDQN}

As mentioned earlier, DESiRED tackles the issue of fixed target delay through the implementation of an intelligent control plane, denoted by the orange box in Figure \ref{fig:iRED-RL}. This intelligent control mechanism is responsible for updating the register that maintains the dynamic target delay threshold, as determined by the DRL decision process. Now, let us provide a comprehensive account of the operational intricacies of the intelligent control plane, elucidating the inputs and outputs in detail.

\begin{figure}[ht]
    \centering
    \includegraphics[width=.8\columnwidth]{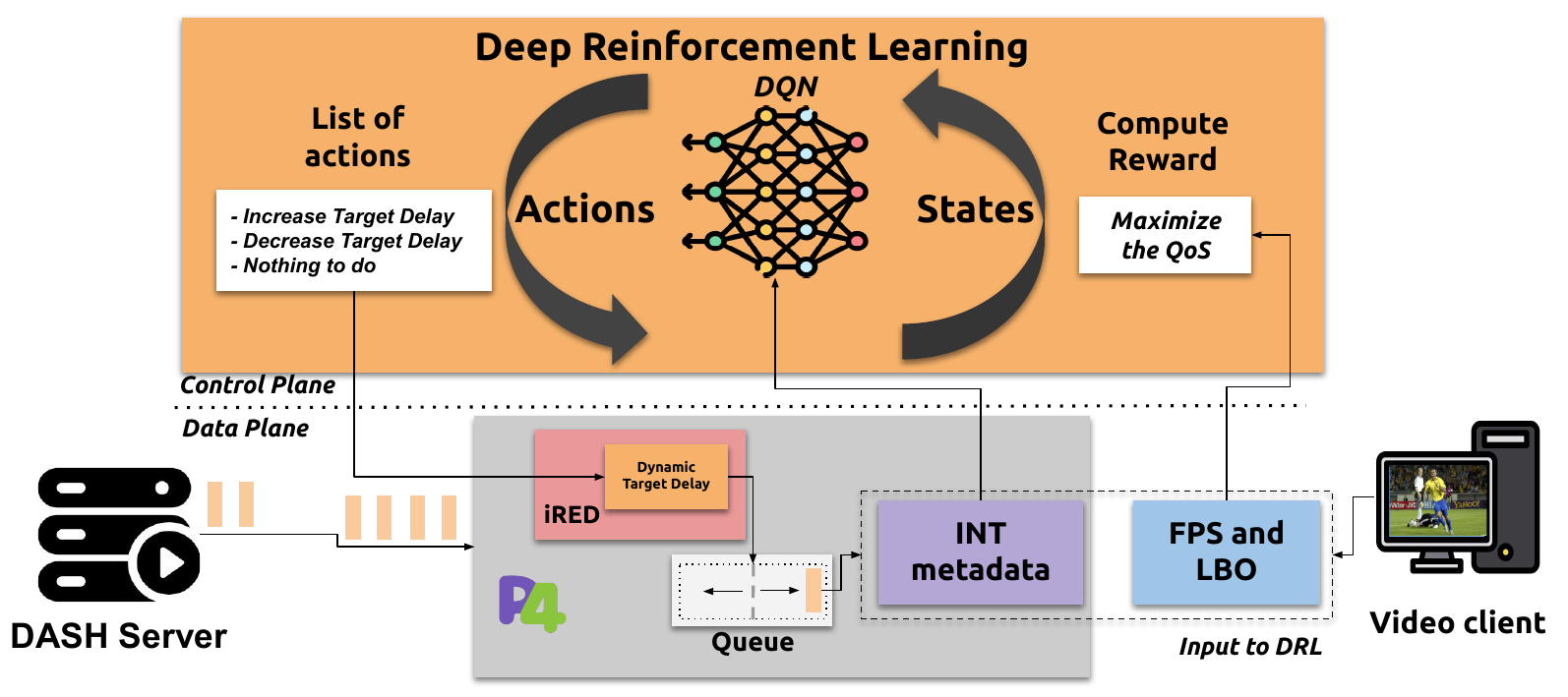}
    \caption{Operation of the Control Plane in DESiRED involves using fine-grained INT measurements as the input layer for the DQN. Additionally, DASH QoS measurements serve as the basis for calculating agent rewards.}
    \label{fig:cl}
\end{figure}

The control plane operates by receiving data from two pivotal sources: the network state and the application state. In this particular implementation, fine-grained INT measurements constitute the input layer for the Deep Q-Network from the network state. The DQN's output layer is responsible for generating the agent's actions. Concurrently, the application state encompasses DASH metrics, including parameters such as FPS and the Local Buffer Occupancy (LBO) of the video player, which play a crucial role in computing the agent's reward. Fig. \ref{fig:cl} illustrates this Control Loop.

INT measurements comprise observations that effectively depict the network's state with remarkable granularity, affording an unprecedented perspective on the extent of congestion. These measurements are acquired within the programmable data plane and subsequently routed to the intelligent control plane. Within the control plane, they are aggregated into compact dataframes, which collectively form what we term the ``observation space." In the context of this study, the term observation space refers to the temporal window within which the intelligent control plane conducts an integrated analysis of both the network's state and the application's behavior.

For each received observation space, the DQN incorporates INT measurements as an input layer. Following neural network processing (refining its internal weights), the DQN generates an action, which is manifested as an activation in one of the neurons within the output layer. In this study, the possible actions include: 1) increasing the target delay; 2) decreasing the target delay; and 3) maintaining the current state (i.e., taking no action).

Subsequently, the control plane retains a record of the executed action and enters a state of anticipation for the forthcoming observation space. Upon the arrival of data from the subsequent observation space, the DRL mechanism evaluates whether the undertaken action has led to the optimization of DASH QoS, particularly with regard to enhancements in FPS and LBO metrics. In the event of a positive outcome, the agent is rewarded, whereas in cases of QoS deterioration, the agent incurs a penalty.

Leveraging insights from the dynamic network traffic patterns, DESiRED demonstrates a remarkable capability to adapt with precision to prevailing congestion conditions. This adaptability facilitates a continuous enhancement in the quality of video services offered.

It is imperative to elucidate that DESiRED is inherently application-agnostic, signifying its capacity to accommodate diverse reward policies tailored to evaluate a wide array of service metrics. This flexibility extends to metrics such as the response time of a web server or even the frame rate in video playback, underscoring its versatility across various service domains.

\section{Evaluation}

In this section, we provide a comprehensive overview of all the components utilized for the thorough evaluation of our proposal. This encompasses a detailed exposition of the research methodology, an in-depth portrayal of the experimental environment and its configuration, the load pattern employed, the DRL mechanism implemented, the metrics and measurements used for comprehensive analysis.

\subsection{Research methodology}

Our methodology is rooted in experimental research aimed at evaluating the effectiveness of the DRL mechanism within DESiRED. Specifically, our objective is to ascertain whether this mechanism can optimize the QoS for MPEG-DASH services by dynamically adapting the target delay under conditions characterized by both stationary and non-stationary loads within a Content Delivery Network (CDN) environment.

In this experiment, our aim is to conduct a comprehensive evaluation of DESiRED in comparison to iRED, where iRED employs fixed target delay settings of 5ms, 20ms, 50ms, and 100ms. We evaluate these approaches under both stationary (low and high) and non-stationary (sinusoidal) load conditions. To mitigate potential biases, each round of the investigation, spanning one hour, was repeated ten times for each approach, resulting in a cumulative duration of over fifty hours across independent runs. Furthermore, to gauge DESiRED's robustness, we aggregated the DRL agents derived from all preceding executions by employing an ensemble approach. This involved combining the model parameters through an exponentially decaying running average, as described by Eq. \ref{eq:polyak_avg} \cite{ensembleRL:92, goodfellow2016deep-ch8}:

\begin{equation}\label{eq:polyak_avg}
    \hat{\theta}^{(t)} = \alpha \hat{\theta}^{(t-1)} + (1 - \alpha) \theta^{(t)}
\end{equation}

\noindent{where $\theta$ represents a parameter from the Q-network; $t$ the gradient descent iterations; $\hat{\theta}^{(t)}$ the average from such parameters ($\frac{1}{t} \Sigma_i \theta^{(i)}$); and $\alpha$ the exponential decaying factor (defined as 2.0).}

%Moreover, we want to evaluate the robustness of the DESiRED, deploying an ensemble model from all previous executions of DRL, using Polyak-Ruppert Averaged \cite{ensembleRL:92} from all models seen toward the end of the training run.

We evaluate the application's performance from the client-side perspective, focusing on three key metrics: FPS LBO, and Rebuffering Rate (Starvation) as measured within the video player. Higher values for FPS and LBO correspond to improved QoS, while for Rebuffering Rate, a lower value signifies enhanced QoS.

In addition to evaluating application quality metrics, we also scrutinize the performance metrics of the DRL agent, encompassing Loss function and Rewards.

\subsection{Environment description}\label{subsec:ired-environment}

\begin{figure}[ht]
    \centering
    \includegraphics[width=1\columnwidth]{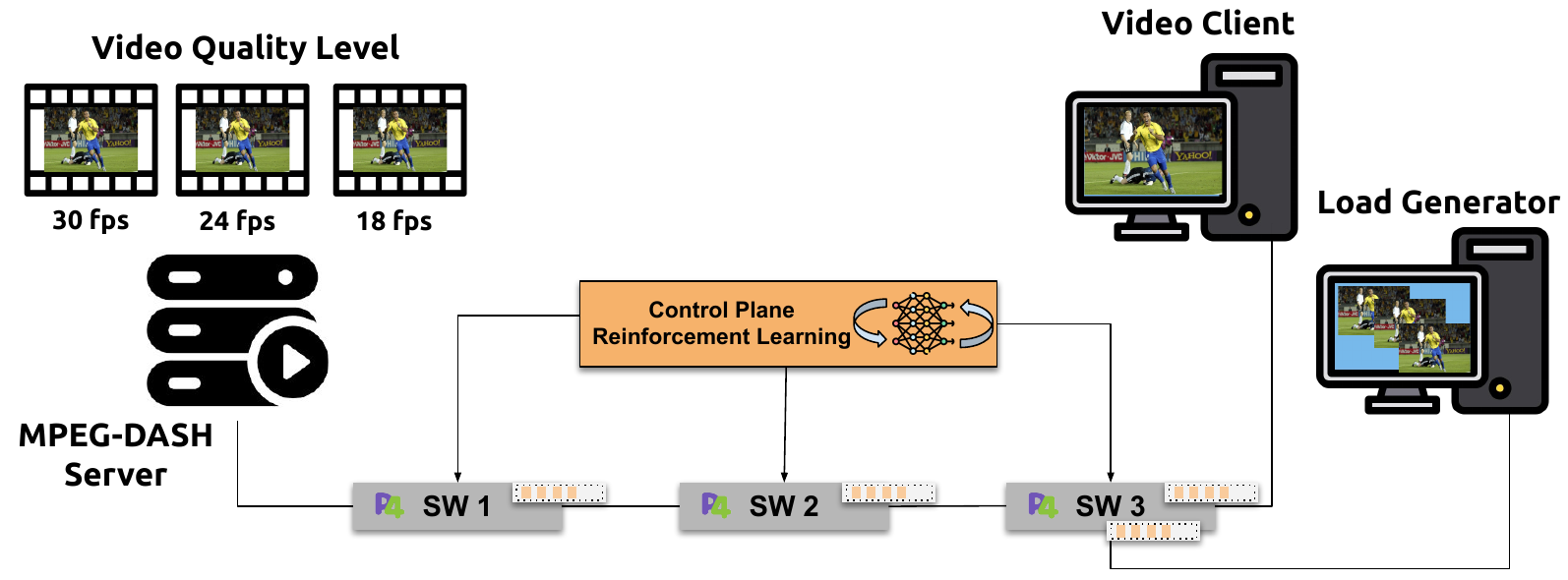}
    \caption{Setup Evaluation.}
    \label{fig:setup}
\end{figure}

The experiment was constructed within a realistic testbed, adopting an Infrastructure as Code (IaC) approach, and implemented using Vagrant, Virtualbox (version 6.1.28), and Ansible (version 2.10.8). In this setup, each infrastructure component is represented by an isolated virtual machine, interlinked through a P4 programmable data plane network, as visually depicted in Figure \ref{fig:setup}.

Each switch in the experiment was equipped with both the iRED and DESiRED approaches. On the control plane side, the DRL engine was implemented, comprising approximately 750 lines of code and utilizing Tensorflow as its backend framework. The CDN was deployed to facilitate an MPEG-DASH service, featuring live streaming of a soccer game and a playlist housing the ten most frequently accessed YouTube videos. Load management was executed using WAVE \cite{Wave:2023} \footnote{https://github.com/ifpb/wave}, a versatile load generator that orchestrates instances of an application over time.

This infrastructure was hosted on a bare-metal server, namely the Dell EMC PowerEdge R720, equipped with 2 Intel Xeon processors (E5-2630 v2, 2.60GHz) boasting 6 cores per socket (amounting to 24 virtual CPUs), 48GB of RAM, a 2TB HDD, and running the Ubuntu 20.04.6 LTS operating system. All pertinent artifacts and resources can be accessed within the repository available at our GitHub\footnote{https://github.com/dcomp-leris/DESiRED.}.

The MPEG-DASH Server serves video content using the DASH standard to both the Video Client and the Load Generator. It offers various configurations, as detailed in Table \ref{tab:dash}, with each configuration having a chunk segment size of 4 seconds. The Video Client dynamically selects and transitions between these configurations based on network traffic conditions and the adaptation logic embedded within the video player.

The infrastructure is equipped with Apache version 2 as the web server, FFmpeg (version 2.8.17) for video encoding, and MP4box (version 0.5.2) for creating the MPEG-DASH manifest files, ensuring seamless video streaming.

\begin{table}[ht] \scriptsize
\centering
\begin{tabular}{ |c|c|c|c|c|c|c| }
\hline
 \textbf{Type} & \textbf{Resolution} & \textbf{FPS} & \textbf{Group of Pictures} & \textbf{Kbps} & \textbf{Buffer} & \textbf{Codec} \\ 
 \hline
 vídeo & 426x240 & 18 & 72 & 280 & 140 & h264  \\
 vídeo & 854x480 & 24 & 96 & 980 & 490 & h264 \\
 vídeo & 1280x720 & 30 & 120 & 2080 & 1040 & h264 \\
 áudio & - & - & - & 128 & - & AAC \\
 áudio & - & - & - & 64 & - & AAC \\
 \hline
 \end{tabular}
 \caption{Video parameters used in an MPEG-DASH Server.}
 \label{tab:dash}
\end{table}

The Video Client utilizes DASH.js, a contemporary DASH reference player equipped with an Adaptive Bitrate Streaming (ABR) algorithm. It employs this ABR algorithm to consume the video stream of the soccer game, with the TCP New Reno congestion control algorithm managing network congestion.

The Load Generator is responsible for introducing network noise, operating the WAVE framework with a variety of loads, including both stationary and non-stationary scenarios. It dynamically adjusts the number of video player instances over time to simulate changing network conditions. Further elaboration on this aspect can be found in Subsection \ref{subsec:load}.

All the switches utilized in this experiment were implemented within the BMv2 software switch environment, incorporating the respective P4 code for both iRED (fixed target delay) and DESiRED (dynamic target delay with DRL) approaches. Across all approaches, telemetry instructions were meticulously programmed to append telemetry metadata to all probe packets. Notably, this experiment follows the out-of-band (ONT) approach, wherein dedicated ONT probes are dispatched from the DASH server to the Video Client. Consequently, no modifications are made to data packets to accommodate telemetry metadata. The specifics of the telemetry metadata, consisting of 32 bytes, gathered at each node within this experiment, are elaborated upon in Table \ref{tab:INTmetadata}.

\begin{table}[ht] \scriptsize
\centering
\begin{tabular}{ |c|c|l| }
\hline
 \textbf{Name} & \textbf{bits} & \textbf{Description} \\ 
 \hline
 Switch ID & 31 &  the switch identification number \\
 Ingress port & 9 & the port number that the packet entered in the switch \\
 Egress port & 9 & the port number that the packet left of the switch \\
 Egress spec & 9 & the port number (Ingress) in which the packet will leave the switch \\
 \makecell{Ingress Global \\ Timestamp} & 48 & the timestamp, in \textmu s, of when the packet entered in the ingress\\
 \makecell{Egress Global \\ Timestamp} & 48 & the timestamp, in \textmu s, of when the packet started processing in the egress\\
 Enq Timestamp & 32 & the timestamp, in \textmu s, of when the packet was enqueue\\
 Enq Qdepth & 19 & the queue depth when the packet was queued\\
 Deq Timedelta & 32 & the time, in \textmu s, that the packet remained in the queue\\
 Deq Qdepth & 19 & the queue depth when the packet was dequeued\\
 \hline
 \end{tabular}
 \caption{INT medatada.}
 \label{tab:INTmetadata}
\end{table}

\subsection{Load Pattern} \label{subsec:load}

The Load Generator, powered by WAVE, orchestrates the instances of video clients over time based on input parameters described by a mathematical function that defines the load pattern. In its current iteration, WAVE supports constant, sinusoidal, and flashcrowd load patterns. It initiates and concludes video player processes, generating network load through genuine video requests (real traffic) that flow from the video player to the MPEG-DASH Server.

\begin{figure}[ht]
    \centering
    \subfigure[Low Load]{
        \includegraphics[width=.47\columnwidth]{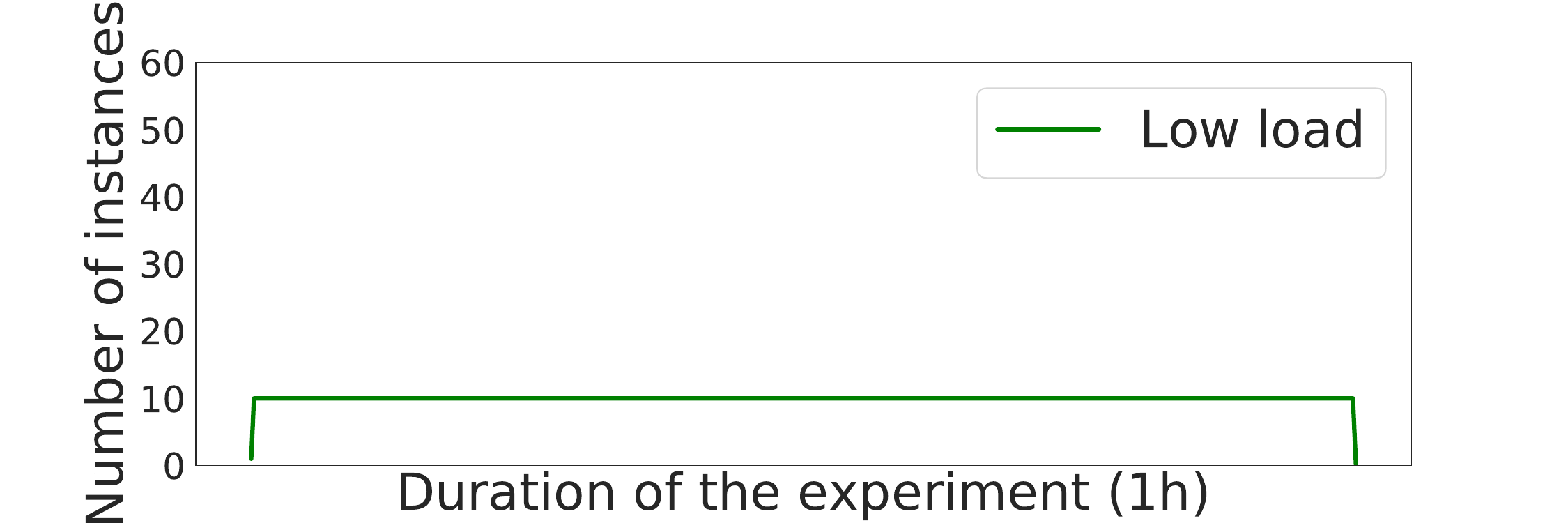}
        \label{fig:Lowload}
    }
    \subfigure[High Load]{
        \includegraphics[width=.47\columnwidth]{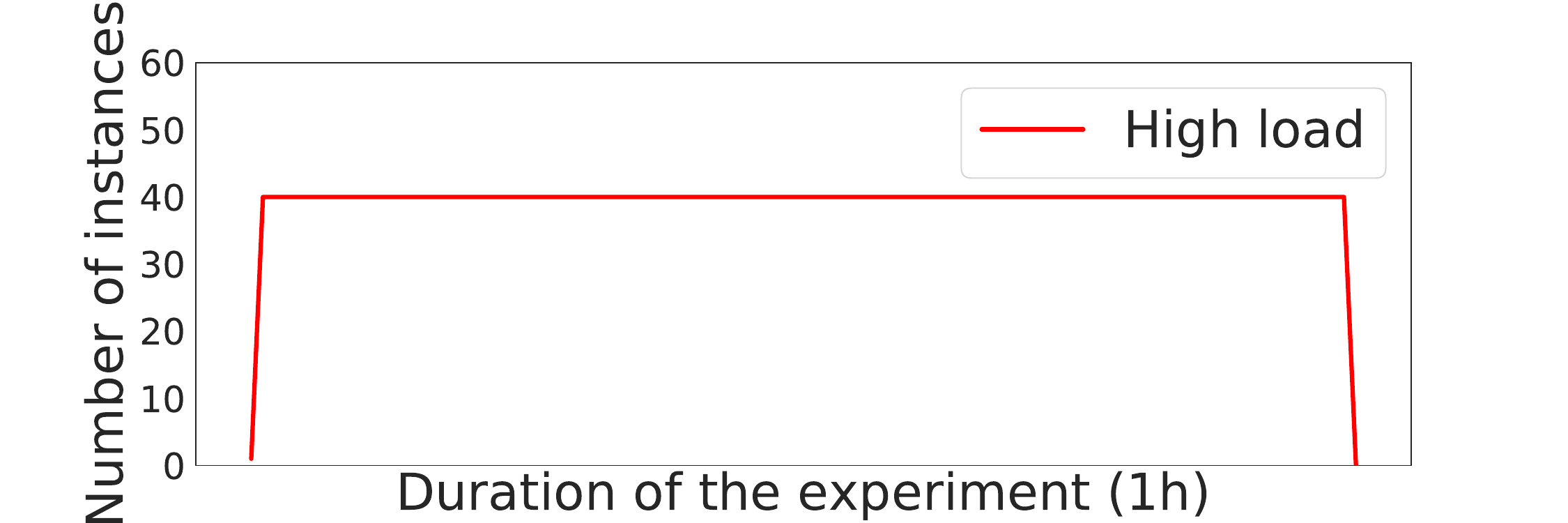}
        \label{fig:Highload}
    }
    \caption{Stationary Loads.}
    \label{fig:stationaryload}
\end{figure}

In this study, our aim is to evaluate DESiRED under various load conditions, aiming to simulate diverse network state scenarios. To achieve this, we employ two distinct categories of load patterns: stationary and non-stationary. For stationary loads, which remain constant throughout the experiment, we classify them into two types: low and high. In this context, a low load is characterized by the presence of ten video client instances operating concurrently throughout the duration of the experiment, as depicted in Figure \ref{fig:Lowload}. Conversely, a high load is characterized by the simultaneous operation of forty video player instances, representing a high-intensity load, as illustrated in Figure \ref{fig:Highload}.

Under low load conditions, it is anticipated that the target delay will be attained relatively infrequently, given the shorter queuing delays that typically prevail. In this scenario, both AQM strategies, whether employing a fixed or dynamic target delay, are likely to yield comparable results in terms of QoS.

However, when the network experiences predominantly high load, the surge in traffic volume can lead to an increase in queue delay, thereby prompting AQM strategies to respond in accordance with the specified target delay, whether fixed or dynamic. In such instances, the dynamic adaptability of DESiRED's target delay is expected to confer advantages in terms of QoS compared to the rigid, fixed target delay approach employed by iRED. This dynamicity enables DESiRED to better accommodate and optimize QoS in the face of fluctuating and demanding network conditions.

\begin{figure}[ht]
    \centering
    \includegraphics[width=.75\columnwidth]{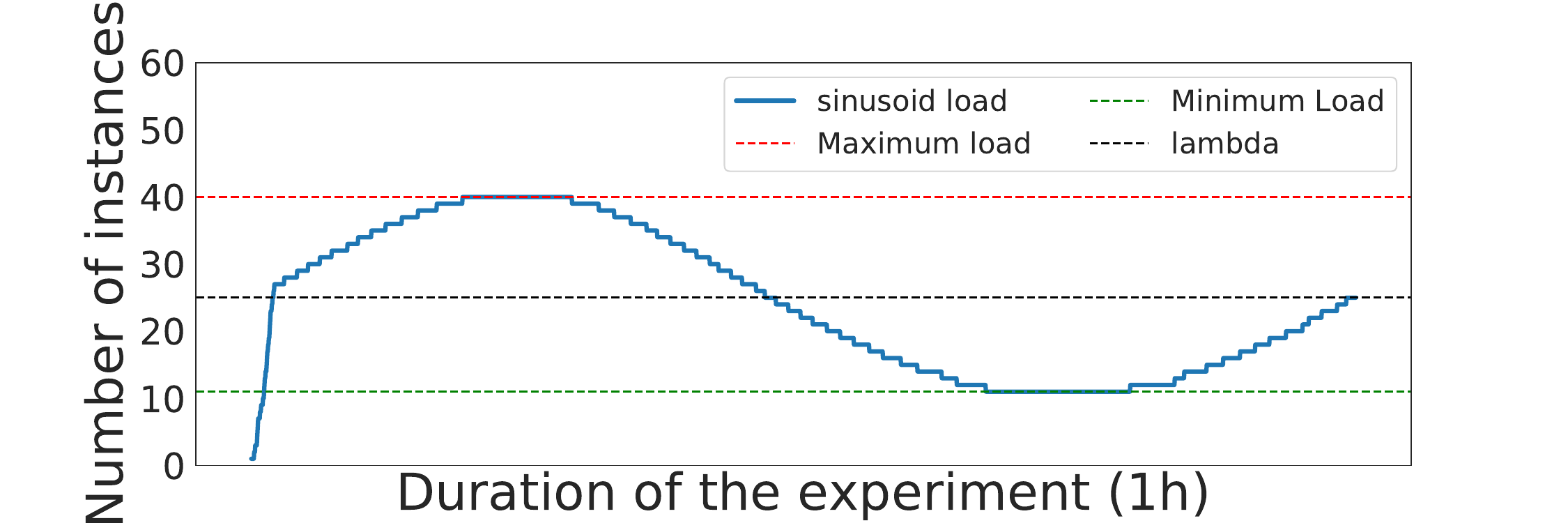}
    \caption{Non-stationary (Sinusoid) Load.}
    \label{fig:sinusoid}
\end{figure}

It is indeed unrealistic to assume that network loads will always remain stationary or static. Consequently, in the second phase of our evaluation, we undertook a more comprehensive evaluation under a realistic load scenario, one that mirrors the dynamic nature of real-world network environments. Our objective was to evaluate non-stationary load patterns, encompassing both peak (high load) and off-peak (low load) periods within a single experiment. To achieve this, we employed a sinusoidal periodic load pattern characterized by the sinusoidal function detailed in Equation \ref{eq:sinusoid}, where \textit{$ A $} represents the amplitude, \textit{$ F $} denotes the frequency, and \textit{$ \lambda $} signifies the phase in radians. The specific input parameters utilized for this evaluation were: $A = 15$, $F = 1$, and $\lambda = 25$, culminating in the load pattern illustrated in Figure \ref{fig:sinusoid}. This approach captures the fluctuations in network load more realistically, offering a dynamic and challenging environment for our evaluation.

\begin{equation}
    \textit{$f(y) = A\sin(F + \lambda)$}    
    \label{eq:sinusoid}
\end{equation}

%In order to evaluate the model in other domains of the problem and thus materialize a type of transfer of learning, we ran WAVE in the experiment 2 with flashcrowd load pattern. The flashcrowd load describes a flash event, that is represented by a large spike or surge in traffic to a particular Web site \cite{flashcrowd}. The flashcrowd is divided into three phases: ramp-up, sustained and ramp-down. Ramp-up is modeled by shock level ($S$), that is an order of magnitude increase in the average request (video clients) rate. Furthermore, it starts in $t_0$ and ends in $t_1$.

%\begin{equation}
%\textit{$rampup = \frac{1}{\log_{10}(1+ S)}$}
%\label{eq:rampup}
%\end{equation}

%Sustained represents the maximum traffic (clients) level at the time interval $t_1$ and $t_2$. It is also modeled by $S$.

%\begin{equation}
%\textit{$sustained = \log_{10}(1+ S)$}
%\label{eq:sustained}
%\end{equation}

%Ramp-down represents the end of the flash event, gradually decreasing the amount of traffic (video clients). In this phase, $n$ is a constant that defines the speed of reduction. Ramp-down is modeled by $n$ and $S$.

%\begin{equation}
%\textit{$rampdown = n\times \log_{10}(1+ S)$}
%\label{eq:rampdown}
%\end{equation}

%\begin{figure}[ht]
%    \centering
%    \includegraphics[width=.9\columnwidth]{fig/flashcrowd.pdf}
%    \caption{Flashcrowd load.}
%    \label{fig:flashcrowd}
%\end{figure}

%In experiment 2, the parameters used were: $Rnorm = 3$, $S = 3$, and $n = 10$. Between waves of flashcrowds, WAVE waits for a time interval ranging from 10 to 100 seconds (random number). 

\subsection{Deep Reinforcement Learning mechanism}\label{subsec:ired-dqn}

%The DRL mechanism was designed in an architecture in which the input layer comprises data provided by network telemetry (INT metadata). In this case, these are fine-grained measurements of the network state. 
%There are two hidden layers with 24 neurons each. The output layer is composed of 7 neurons, in which each neuron represents an action, as detailed in Fig \ref{fig:dqnarch}.

\begin{figure}[ht]
    \centering
    \includegraphics[width=.8\columnwidth]{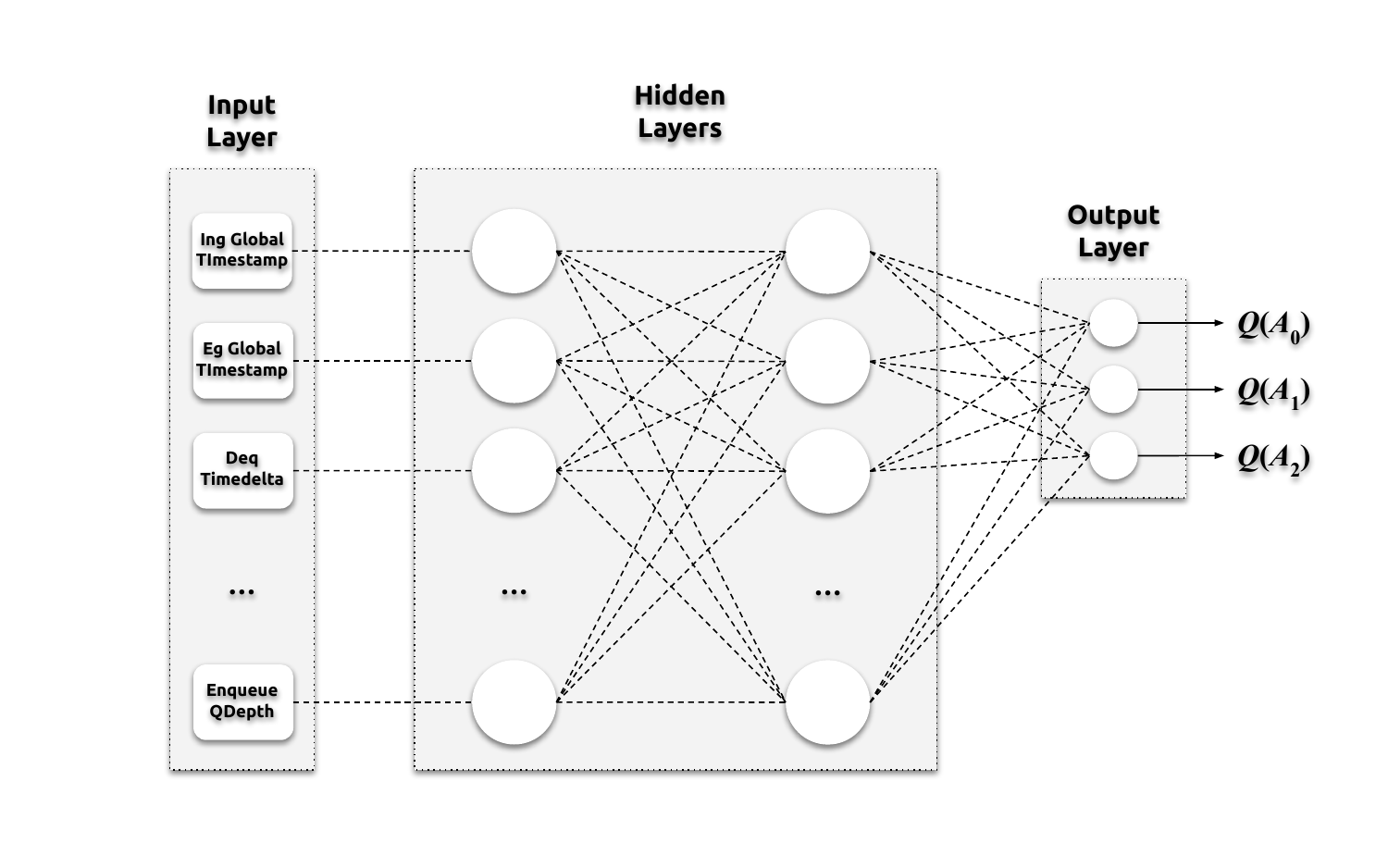}
    \caption{DESiRED DQN architecture. The input layer is a network of fine-grained measurements, provided by INT. Hidden Layers make up the DQN. The actions are defined in the Output Layer.}
    \label{fig:dqnarch}
\end{figure}

%modified dqn architecture
To accomplish the objectives outlined in this paper, we tailored the DQN architecture and agent-environment workflow to align with the distinctive characteristics of the DESiRED environment, as elucidated in Subsection \ref{subsec:ired-environment}. In doing so, we designed the DQN using a Multi-layer Perceptron (MLP) architecture, which is well-suited for handling the tabular nature of network telemetry metadata. The MLP network adopted in our approach consists of an input layer featuring units corresponding to each INT feature, two hidden layers each comprising 24 neurons, and an output layer containing units for each possible action that the agent can undertake, as depicted in Figure \ref{fig:dqnarch}. Importantly, both the online and target networks share this identical architecture. Table \ref{tab:dqn_hyperparameters} provides a detailed breakdown of the hyperparameters utilized for training DESiRED.

%next steps:
%--> create a table with all hyperparameters of the DQN
%--> describe the modified agent-environment interaction cycle

%The actions are described in Table \ref{tab:actions}. The reason for using the increment twice more significant than the decrement is that we want to encourage the agent to quickly accommodate transient congestion, while still offering an option to decrease the target delay when necessary, in a similar way to that discussed in \cite{QTCP:2019}.

%\begin{table}[ht]
%\scriptsize
%\centering
%\begin{tabular}{ |c|c|l| }
%\hline
% \textbf{Action Number} & \textbf{Value} & \textbf{Description} \\ 
% \hline
% 0 & + 2ms &  increase target delay in switch 3 \\
% 1 & - 1ms &  decrease target delay in switch 3 \\
% 2 & + 2ms & increase target delay in switch 2  \\
% 3 & - 1ms & decrease target delay in switch 2 \\
% 4 & + 2ms & increase target delay in switch 1 \\
% 5 & - 1ms & decrease target delay in switch 1\\
% 6 & - & Nothing\\
% \hline
% \end{tabular}
% \caption{Actions.}
% \label{tab:actions}
%\end{table}

\begin{table}[ht]
    \begin{threeparttable}
        \scriptsize
        \centering
            \begin{tabular}{ |l|c|l| }
                \hline
                \multicolumn{1}{|c|}{\textbf{Hyperparameter}} & \textbf{Value} & \multicolumn{1}{|c|}{\textbf{Description}} \\ 
                \hline
                    Q-network input layer dimension  & 19 & A scalar defining the state input shape. \\
                    Q-network hidden layers &  2 & A scalar defining the Q-network depth. \\
                    Q-network hidden units  & 24 & A scalar defining the Q-network non-linear computing units. \\
                    Q-network output layer dimension  & 3 & A scalar defining the Q-network predictions output shape. \\
                    Hidden units activation function & ReLU\tnote{1} & The non-linear activation function computed by hidden units. \\
                    Output units activation function & Linear & The activation function computed by the output layer. \\
                    Optimization function & SGD\tnote{2} & \makecell[l]{The function used to adjust the Q-network weights in order to \\ minimize the predictions error in relation to the expected output.} \\
                    SGD momentum & 0.9 & \makecell[l]{A scalar defining the momentum included in the \\ optimization equation to accelerate the gradient descent.} \\
                    Learning rate & 1e-3 & A scalar determining the pace at which the weights are updated. \\
                    Loss function & MSE\tnote{3} & The function used to compute the Bellman equation error. \\
                    Gamma & 0.99 & A scalar determining the discount factor in the Q-Learning update. \\
                    Tau & 1e4 & \makecell[l]{A scalar value determining how many updates the online network \\ should perform before updating the target network \\ (it corresponds to the parameter $C$ depicted in the Fig. \ref{fig:rl_dqn-double_q-learning}).} \\
                    Experience replay capacity & 1e6 & \makecell[l]{A scalar defining the size of the list in which the agent's \\ experience will be stored.} \\
                    Experience replay minimum memory & 100 & \makecell[l]{A scalar defining the minimum experiences that should be \\ stored before updating the online network.} \\
                    Mini batch size & 32 & \makecell[l]{A scalar defining the number of experience samples over which \\ the Q-network will be updated.} \\
                    Starting epsilon & 1.0 & \makecell[l]{A scalar defining the initial probability to take random actions \\ in the $\epsilon$-greedy exploration.} \\
                    Ending epsilon & 0.01 & \makecell[l]{A scalar defining the final probability to take random actions \\ in the $\epsilon$-greedy exploration.} \\
                    Epsilon decay steps & 250 & \makecell[l]{A scalar determining how many steps the probability to \\ take random actions in the $\epsilon$-greedy exploration should decrease \\ linearly before the exponential decay.} \\
                    Epsilon exponential decay & 0.99 & \makecell[l]{A scalar determining exponential decay of the probability to take \\ random actions in the $\epsilon$-greedy exploration.} \\ 
                \hline
            \end{tabular}
            \begin{tablenotes}
                \footnotesize{\item [1] Rectified Linear Unit.}
                \footnotesize{\item [2] Stochastic Gradient Descent with Nesterov Momentum.}
                \footnotesize{\item [3] Mean Squared Error.}
            \end{tablenotes}
    \end{threeparttable}
    \caption{DQN hyperparameters.}\label{tab:dqn_hyperparameters}
\end{table}

\begin{table}[ht]
\scriptsize
\centering
\begin{tabular}{ |c|c|l| }
\hline
 \textbf{Action Number} & \textbf{Value} & \textbf{Description} \\ 
 \hline
 0 & + 2ms &  increase target delay in all switches until 70ms (upper limit) \\
 1 & - 1ms &  decrease target delay in all switches until 20ms (lower limit)\\
 2 & - & do nothing \\
 \hline
 \end{tabular}
 \caption{DESiRED actions space.}
 \label{tab:actions}
\end{table}

%modified agent-environment interaction
To facilitate the desired agent-environment interaction, we formulated the agent's behavior as an MDP with the video chunk size serving as the discrete time steps. In this framework, DESiRED operates within the environment, dynamically adjusting the target delay in all switches at 4-second intervals, synchronized with the video chunk size. A comprehensive discussion regarding the strategy of simultaneous actuation in all switches versus individual actuation in each switch is presented in Section \ref{sec:lessons}. The agent's action space is delineated in Table \ref{tab:actions}, where it is evident that the action to increase the target delay brings about a modification that is proportionally twice as substantial as the decrease action. This choice was made to prompt DESiRED to respond promptly to transient congestion while retaining the flexibility to decrease the target delay when necessary, mirroring the rationale discussed in \cite{QTCP:2019}.

%Nonetheless, it should be noted that the reward could not be calculated immediately after the action taken in the current state, since the consequence of such action would only be noticeable in the next state (video chunk). This phenomenon occurs because both TCP and the ABR algorithm have control mechanisms to alleviate the congestion on the network, hence, we delayed the reward calculation until the next state observation.

It's important to highlight that the calculation of rewards does not occur immediately after an action is taken in the current state. This delay in reward calculation is attributed to the fact that the effects of the agent's action do not manifest instantly, primarily due to the inherent control mechanisms incorporated within TCP and ABR systems, as detailed in \cite{dashjs:2018}. Consequently, the computation of rewards is deferred until the subsequent state's observation. In this context, the agent relies on network status data derived from INT measurements to form its states, selects actions, and is subsequently rewarded based on its ability to optimize the video's QoS, which is characterized by metrics such as FPS and LBO.

%Although, it is worth mentioning that the reward could not be calculated immediately after the action taken, since the effect of such action does not occur instantly. This phenomenon is due to the control mechanisms that both TCP and ABR provide to alleviate network congestion, hence, we delayed the reward calculation until the next state's observation. In this context, the agent receives network status data from INT measurements as states, takes actions, and is rewarded if it manages to maximize the video's QoS; which is characterized by the FPS (Frames Per Second) and LBO (Local Buffer Occupancy).

Indeed, the intrinsic correlation between metrics such as LBO and FPS presents a challenge when devising a reward policy. As the LBO increases, there is a tendency for the FPS to also increase. However, this relationship is not always straightforward due to the complex dynamics of network congestion and video streaming.

To calculate a reward ($R_{t+1}$) for a specific action ($A_t$), we adopt a strategy that first evaluates whether the LBO in the next state ($LBO_{t+1}$) improves compared to the LBO observed when the action was executed ($LBO_t$). Subsequently, a reward score is assigned based on the effects of this action on both the next state's LBO and FPS ($FPS_{t+1}$). Consequently, the agent receives maximum reward whenever the action taken leads to the maximization of $LBO_{t+1}$, and is penalized in an inversely proportional manner if the video experiences stalls. The algorithmic logic for calculating rewards is detailed in Algorithm \ref{alg:reward}. This approach ensures that the agent's reward is contingent on its capacity to optimize both LBO and FPS, balancing the trade-offs inherent to video streaming in dynamic network conditions.

\begin{algorithm}
	\caption{DESiRED reward policy algorithm.}\label{alg:reward}
	\begin{algorithmic}[1]
		\Function{$calculate\_reward$}{$LBO_t$, $LBO_{t+1}$, $FPS_t$, $FPS_{t+1}$}
            \If{$LBO_{t+1} > LBO_t$}
                \If{$LBO_{t+1} > 30$}
                    \State $R_{t+1} \gets 2$
                \ElsIf{$LBO_{t+1} < 30$}
                    \If{$FPS_{t+1} == 30$}
                        \State $R_{t+1} \gets 1$
                    \ElsIf{$FPS_{t+1} == 24$}
                        \State $R_{t+1} \gets 0.5$
                    \Else
                        \State $R_{t+1} \gets 0.1$
                    \EndIf
                \EndIf
            \EndIf
            \\
            \If{$LBO_{t+1} < LBO_t$}
                \If{$LBO_{t+1} > 30$}
                    \State $R_{t+1} \gets 2$
                \ElsIf{$LBO_{t+1} < 30$}
                    \If{$FPS_{t+1} == 30$}
                        \State $R_{t+1} \gets 1$
                    \ElsIf{$FPS_{t+1} == 24$}
                        \State $R_{t+1} \gets 0.5$
                    \Else
                        \State $R_{t+1} \gets -2$
                    \EndIf
                \EndIf
            \EndIf
        \EndFunction
	\end{algorithmic} 
\end{algorithm}

These actions were executed according to the $\epsilon$-greedy strategy as elucidated in Subsection \ref{subsubsec:dqn}. To implement this strategy, we established initial and final probabilities for taking random actions, specified the number of decaying steps, and defined an exponential decay factor (as outlined in Table \ref{tab:dqn_hyperparameters}). In this scheme, $\epsilon$ commences its linear decrease over a span of 250 time steps to facilitate exploration. Subsequently, the probability of selecting random actions is exponentially reduced, gradually transitioning to a minimal value to emphasize exploitation over exploration. This strategy allows the agent to strike a balance between exploring new actions and exploiting its existing knowledge as it interacts with the environment.

Taking into consideration the agent's action frequency of once every 4 seconds and the requirement for 250 iterations to initiate the exponential decay of $\epsilon$, the exploration phase is expected to persist for approximately 17 minutes (equivalent to 1000 seconds). In tandem, the experience replay memory buffer necessitates a minimum of 100 samples to facilitate the online network parameter updates (as indicated in Table \ref{tab:dqn_hyperparameters}. Since experiences resulting from the agent-environment interaction are stored every 8 seconds, it would take approximately 13 minutes (or 800 seconds) for this condition to be met. Consequently, the online network undergoes an update each time a new experience is stored, as illustrated in Figures \ref{fig:rl_dqn-env_interaction} and \ref{fig:rl_dqn-double_q-learning}.

In the case of the non-stationary load, it follows a trajectory of 15 minutes to reach its peak, maintains a plateau for an additional 15 minutes, and subsequently begins to decline. During this period, the agent explores the action space during the ascending phase of the sinusoidal curve and exploits these actions during the plateau and descending phases. Consequently, when the exploitation stage commences, the agent should have already gleaned insights from past experiences, encompassing both low and high load scenarios. This enables the agent to adapt and respond effectively to the fluctuating network conditions.

%Such actions were performed following the $\epsilon$-greedy strategy described in the subsubsec. \ref{subsubsec:dqn}. For such, we defined a initial probability to take random actions that decreases linearly throughout 250 time steps, in order to favor exploration. Then, this probability starts decreasing exponentially afterwards in order to enable exploitation

%The reward policy was built to incentive the agent to maximize the video's QoS, observing video metrics FPS (Frames Per Second) and LBO (Local Buffer Occupancy), as shown in Table \ref{tab:reward}. That is, in this evaluation, the agent receives network status data from INT measurements, takes actions, and is rewarded if it manages to maximize the video quality of service metrics.

%\begin{table}[ht]
%\scriptsize
%\centering
%\begin{tabular}{ |c|c| }
%\hline
% \textbf{Reward value} & \textbf{Condition} \\ 
% \hline
% + 2  & FPS = 30 and LBO $>$ 30\\
% + 1.8 &  FPS = 30 and LBO $<$ 30\\
% + 1.5  & FPS = 24 and LBO $>$ 20\\
% + 1.3  & FPS = 24 and LBO $<$ 20\\
%- 0.1 & All other conditions (FPS = 18)\\
% \hline
% \end{tabular}
% \caption{Reward policy.}
% \label{tab:reward}
%\end{table}

\subsection{Metrics and Measurements}

On the video client side, we evaluate the QoS by monitoring key metrics, including:
\begin{itemize}
    \item FPS (Frames Per Second): This metric quantifies the number of frames displayed per second on the screen, reflecting the smoothness of the video playback.
    \item LBO (Local Buffer Occupancy): LBO measures the remaining time, in seconds, for frames stored in the player's local buffer. It provides insights into the buffer's capacity to absorb network fluctuations and maintain continuous playback.
\end{itemize}

From these primary metrics, we derive additional insights, including:
\begin{itemize}
    \item Resolution Distribution: We analyze the percentage of video content played at different resolutions (Maximum, Medium, and Minimum) to assess the adaptive streaming capabilities.
    \item Rebuffering Rate: This metric represents the percentage of time during which the video experiences stalls or freezes on the screen, indicating interruptions in playback.
\end{itemize}

To facilitate these measurements, we configure the DASH.js player to log these metrics on a per-second basis. Within the DRL mechanism, we focus on evaluating the performance metrics of the DQN:

\begin{itemize}
    \item Loss: This metric is calculated as the Mean Squared Error (MSE) between the predicted q-values for the current and next states. It reflects the convergence and accuracy of the DQN's predictions.
    \item Reward: Reward represents the cumulative rewards and penalties acquired throughout the experiment. It offers insights into the agent's performance in maximizing QoS.
\end{itemize}

Additionally, we capture the action history for each experiment, documenting the agent's selected actions at each observation space (every 4 seconds). These metrics provide a comprehensive view of the agent's learning and adaptation throughout the experiment.

%At the data plane (P4 code), we use counters to save the \textbf{number of drops} (performed by the AQM algorithm) for each approach. These counters are positioned at the Ingress block and incremented in each drop invocation. 

\section{Results}

In this section, we will present the outcomes of our experiments, where we evaluate how DESiRED enhances the QoS of MPEG-DASH. We offer an in-depth analysis from the client-side perspective, showcasing the results and delving into instances where video QoS has benefited from the dynamic adjustments facilitated by DESiRED. Furthermore, we scrutinize the performance of the DRL model, presenting evidence that the agent has successfully learned the designated policy and has been able to identify an optimal target delay value that maximizes QoS across the range of experiments conducted.

\subsection{Stationary Loads}

The motivation behind evaluating performance under stationary loads stemmed from the necessity to ascertain whether the DRL agent would exhibit distinct learning behaviors during moments of low load (ample resources) and high load (congested resources) across separate executions.

\begin{figure}[ht]
    \centering
    \subfigure[FPS]{
        \includegraphics[width=.45\columnwidth]{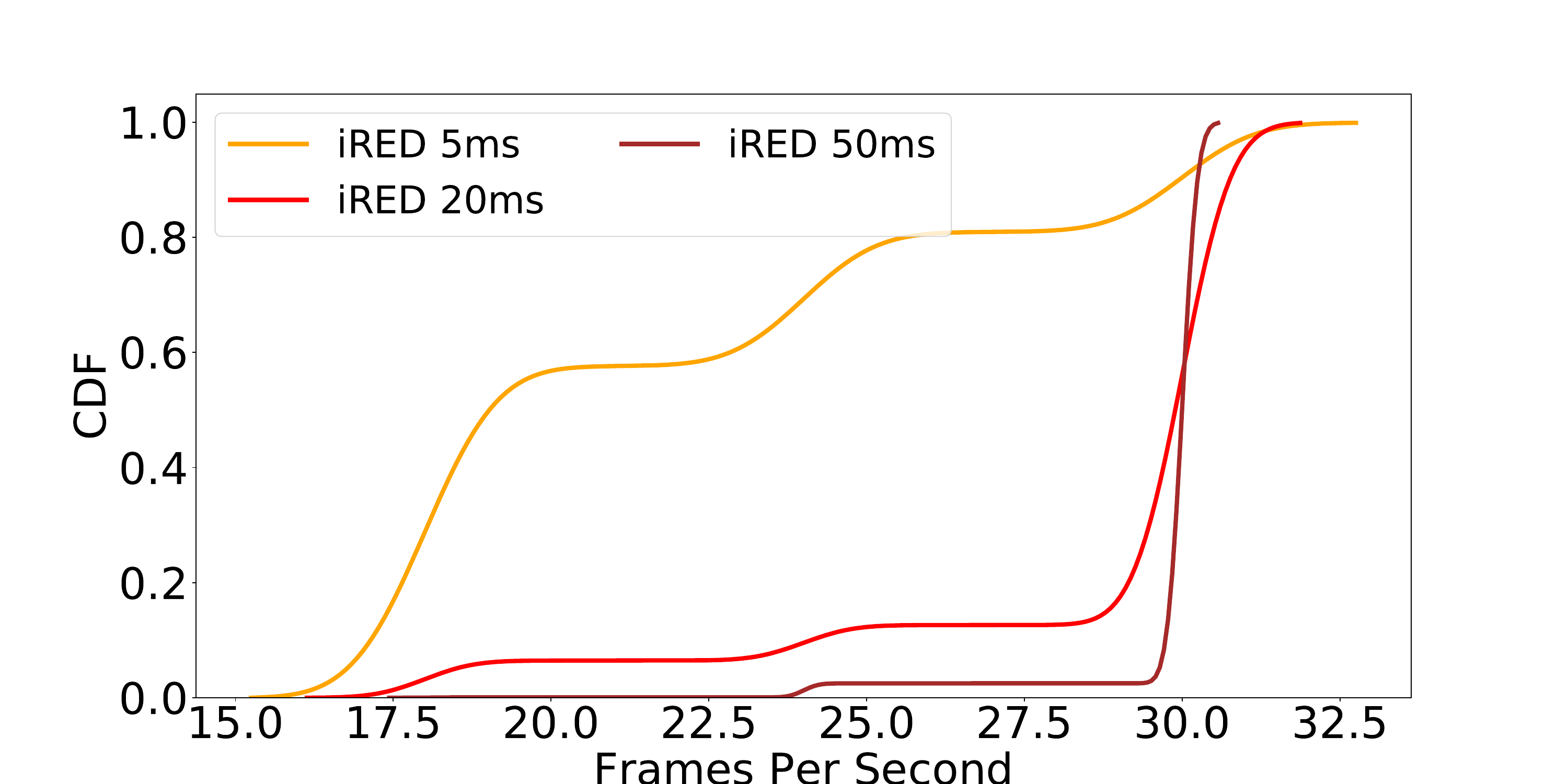}
        \label{fig:fps_lowload}
    }
    \subfigure[LBO]{
        \includegraphics[width=.45\columnwidth]{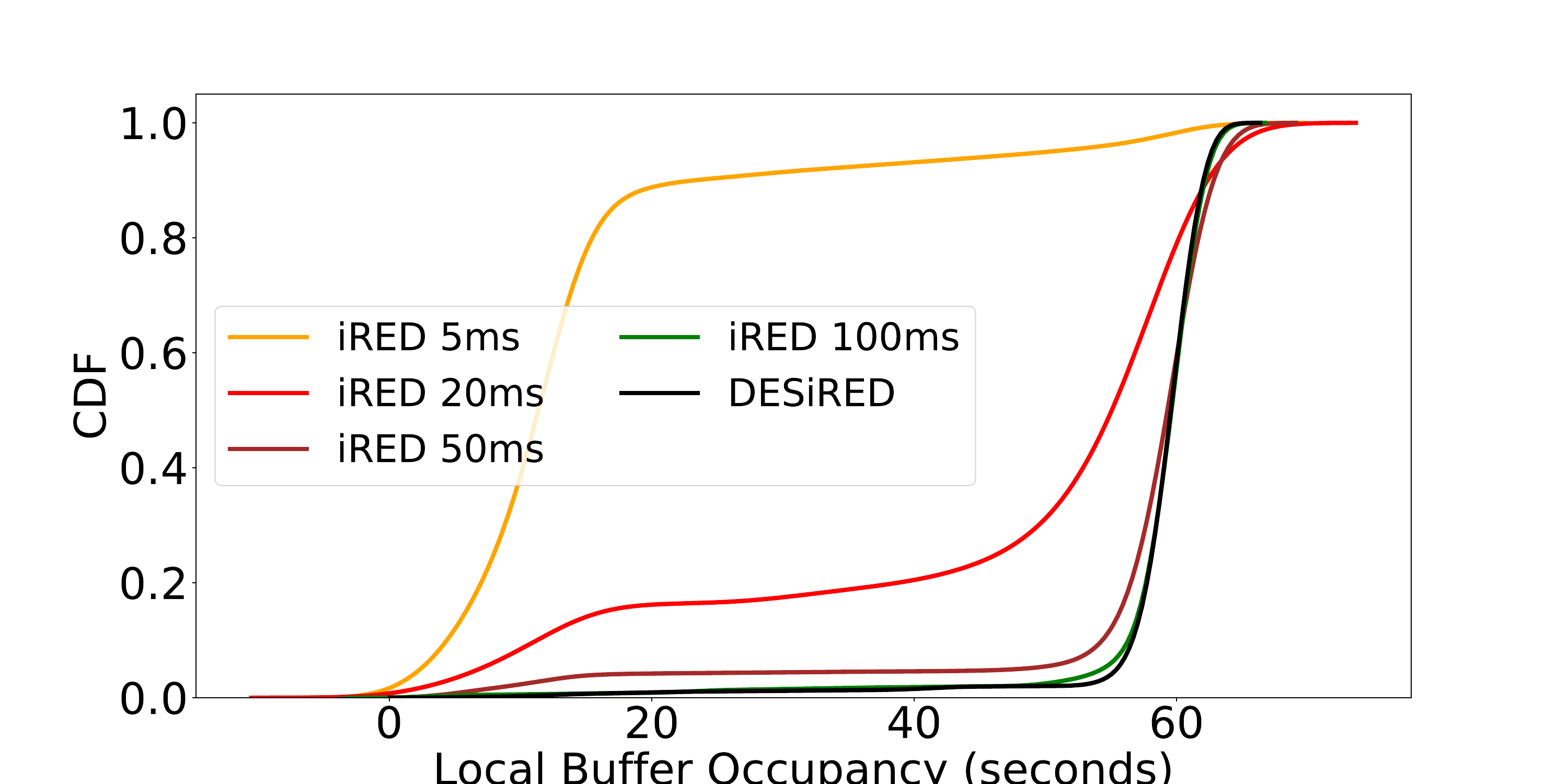}
        \label{fig:lbo_lowload}
    }
    \caption{Low Load - Characterized by only ten video player instances managed by WAVE (Load Generator).}
    \label{fig:lowload}
\end{figure}

When the network load predominantly remains low, as illustrated in Figure \ref{fig:Lowload}, network resources are readily available. In such scenarios, there is minimal contention for the use of the queue, resulting in limited or no intervention from auxiliary congestion control mechanisms like AQM. This phenomenon can be observed from the perspective of the video client, as depicted in Figure \ref{fig:lowload}.

Figures \ref{fig:fps_lowload} and \ref{fig:lbo_lowload} illustrate the Cumulative Distribution Function (CDF) of FPS and LBO under low load conditions. In Figure \ref{fig:fps_lowload}, we observe some variation in FPS for iRED with a fixed target delay of 5ms and 20ms/50ms. Conversely, in the cases of iRED with a fixed target delay of 100ms and DESiRED, the video client consistently played the video at 30 FPS throughout all experiments.

Concerning LBO, as depicted in Figure \ref{fig:lbo_lowload}, the results exhibit similar behavior across approaches, with the local buffer maintaining a near-full state for most of the evaluations, approximately 60 seconds. The only exception is the iRED with a 5ms fixed target delay. In this specific scenario, the use of such a small threshold value appears to have triggered a higher frequency of AQM actions. This, in turn, might have led to more frequent drops within a time interval of less than one Round-Trip Time (RTT), as discussed in \cite{Bufferbloat:2011}. Paradoxically, this increased AQM activity, rather than alleviating congestion, may have exacerbated the situation, demonstrating the potential for unintended side effects when setting overly aggressive congestion control thresholds.

\begin{figure}[ht]
    \centering
    \subfigure[FPS]{
        \includegraphics[width=.45\columnwidth]{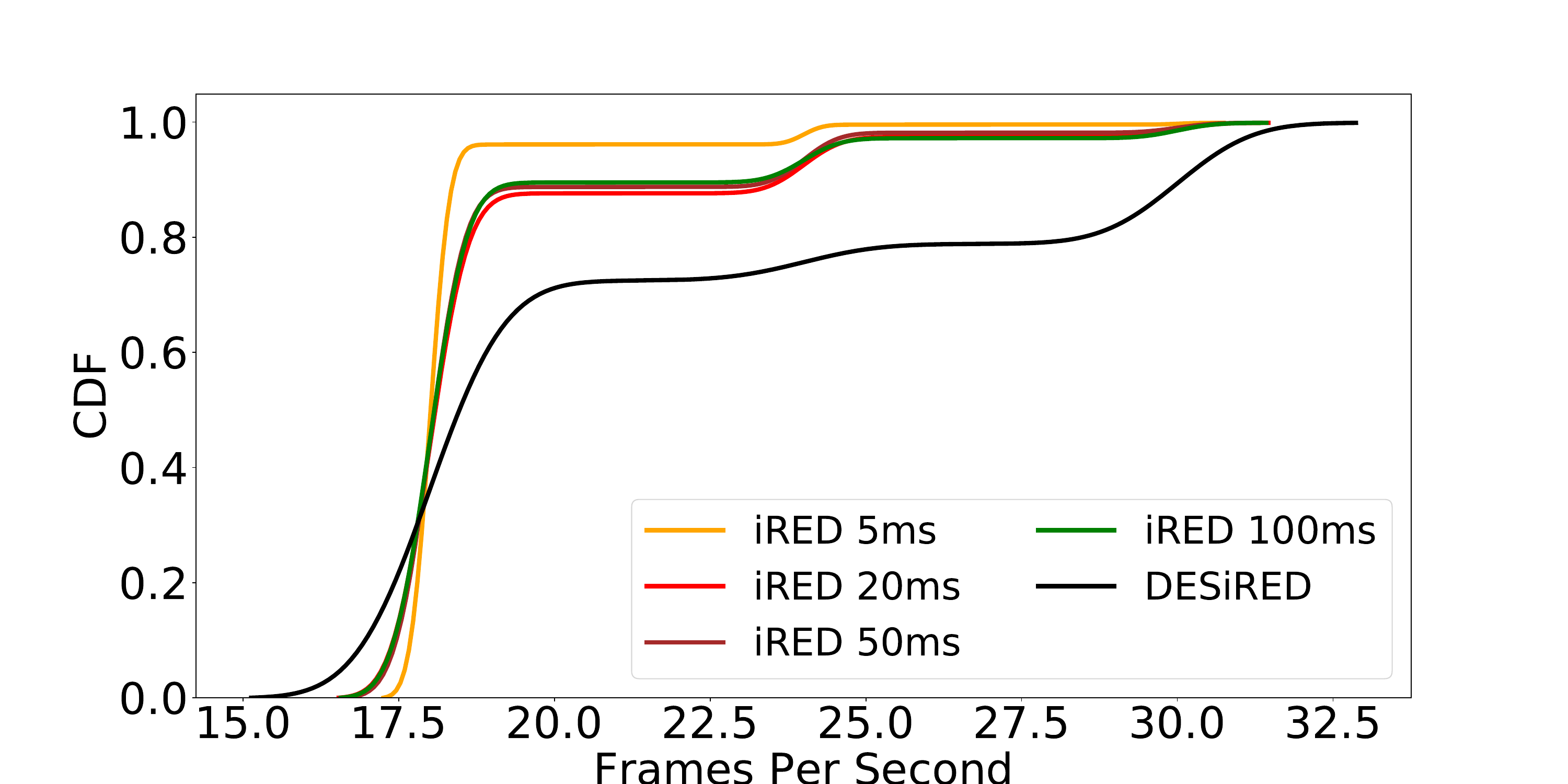}
        \label{fig:CDF_FPS_HighLoad}
    }
    \subfigure[LBO]{
        \includegraphics[width=.45\columnwidth]{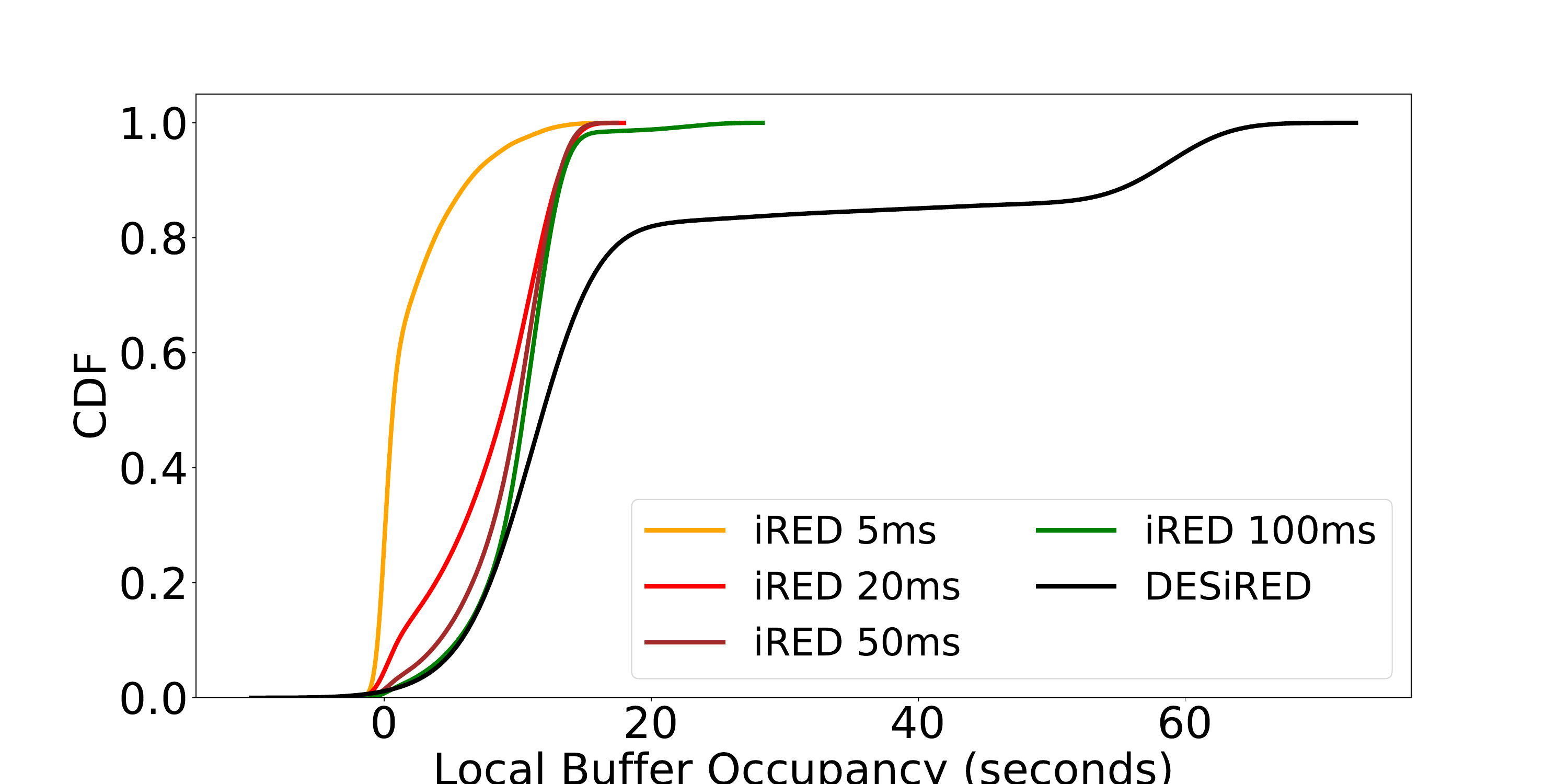}
        \label{fig:CDF_LBO_HighLoad}
    }
    \caption{High Load. Characterized by forty video player instances managed by WAVE (Load Generator).}
    \label{fig:highload}
\end{figure}

Conversely, when the network experiences predominantly high load conditions, as illustrated in Figure \ref{fig:Highload}, the dynamics shift significantly. In such scenarios, all approaches employing fixed target delay mechanisms encounter challenges in maintaining acceptable MPEG-DASH QoS. DESiRED, on the other hand, manages to distinguish itself from the fixed target delay approaches, as evident in Figure \ref{fig:highload}.

To gain a deeper understanding of these results, it's important to clarify some aspects of the Adaptive Bitrate (ABR) adaptation logic employed by the DASH.js player, as described in \cite{dashjs:2018}. The adaptation logic used in DASH.js, known as DYNAMIC, employs two different algorithms at different stages of video playback. During instances when buffer levels (LBO) are low, such as startup and seek events, a straightforward THROUGHPUT algorithm (based on throughput) is utilized. Conversely, when buffer levels are high, the player switches to the BOLA algorithm \cite{BOLA:2020}. This dynamic adaptation approach aims to optimize video streaming under varying network conditions, aligning the bitrate selection algorithm with the network's congestion state.

DYNAMIC starts with THROUGHPUT until the buffer level reaches 10s or more. From this point on, DYNAMIC switches to BOLA which chooses a bitrate at least as high as the bitrate chosen by THROUGHPUT. DYNAMIC switches back to THROUGHPUT when the buffer level falls below 10s and BOLA chooses a bitrate lower than THROUGHPUT \cite{dashjs:2018}.

Indeed, from the perspective of the video player's adaptation logic, the LBO metric proves to be far more sensitive to variations in network buffer levels compared to FPS. It's important to note that changes in bitrate and FPS should only occur when the LBO drops below 10 seconds. Consequently, it is logical to aim for maintaining an LBO greater than 10 seconds for the majority of the time, as this instructs the ABR algorithm to select the highest-quality video levels.

Figure \ref{fig:CDF_LBO_HighLoad}, which pertains to LBO, contributes significantly to understanding why DESiRED achieves superior FPS levels, as indicated in Figure \ref{fig:CDF_FPS_HighLoad}. In this context, it is plausible to surmise that fine-tuning the target delay has provided an advantage in terms of preserving a sufficient LBO during periods of severe network congestion. This, in turn, aids the ABR algorithm in making optimal bitrate and quality level selections, ultimately leading to improved video QoS.

\subsection{Non-stationary Load} \label{sec:stationaryLoad}

Recognizing the dynamic nature of network traffic, we embarked on an evaluation under non-stationary load conditions. To achieve this, we leveraged the WAVE framework, which effectively managed the execution of video client instances over time, adhering to a mathematical model of sinusoidal periodic load as detailed in Subsection \ref{subsec:load}.

The choice of a sinusoidal periodic load model holds significance because it encapsulates moments of congestion and resource relief in the network, particularly within router buffers, within a single execution. This approach allows us to evaluate our agent's performance in situations of both high congestion, where rapid adaptation is crucial, and congestion-free states where shared resources are not overwhelmed. In essence, our expectation is that the agent will learn distinct patterns that differentiate between these varying states.

This evaluation under non-stationary load conditions provides valuable insights into how the agent responds to fluctuations in network congestion, thereby contributing to a more comprehensive understanding of its adaptability and effectiveness.

\begin{figure}[ht]
    \centering
    \includegraphics[width=.5\columnwidth]{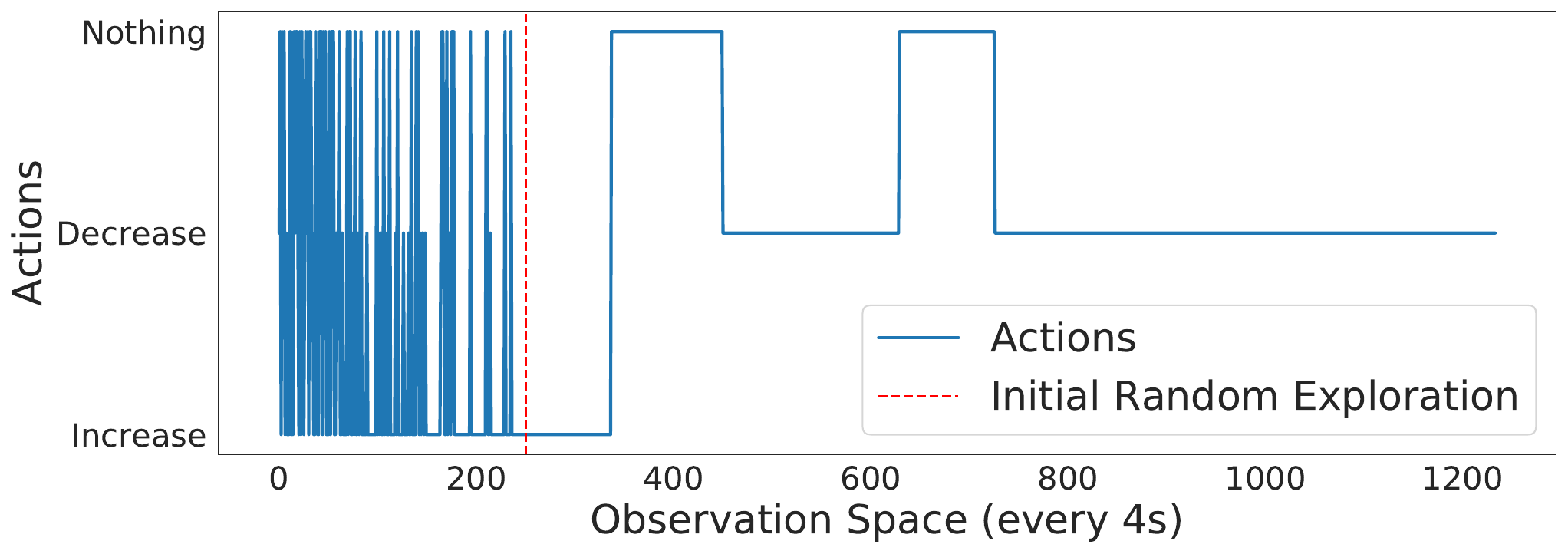}
    \caption{Actions performed by the agent in the environment. After the initial random exploration, the agent finds the best target delay value to maximize the QoS of MPEG-DASH.}
    \label{fig:actions}
\end{figure}

The initial result we would like to present pertains to the actions taken by the agent (DESiRED) within the network environment. Figure \ref{fig:actions} provides an overview of the agent's actions throughout the experiment. Notably, there is an initial phase of random exploration (indicated by the vertical dashed red line) extending up to the first 250 observations. During this exploratory phase, the agent gathers data about the network state, which is used to populate the experience replay buffer (as outlined in Subsection \ref{subsubsec:dqn}).

Subsequent to this initial exploration phase, the agent commences taking actions based on its learned knowledge, drawing from the experiences stored in the experience replay buffer. It's important to highlight that this buffer is continually updated, enabling the agent to learn from new states. Consequently, the agent can adapt to previously unseen states, a capability that proves particularly valuable in scenarios with non-stationary loads.

This analysis of the agent's actions provides insights into its learning process and the transition from exploration to exploitation as it becomes more knowledgeable about the environment.

Analyzing the agent's actions, it becomes apparent that during the initial phase of the experiment, characterized by an increase in network load, the agent frequently opted to increase the value of the target delay. Subsequently, as the load stabilized, the agent chose to take no action, potentially reducing the overhead of control plane operations in the data plane. Towards the end of the experiment, as the network load decreased, the agent shifted its strategy towards reducing the target delay.

\begin{figure}[ht]
    \centering
    \subfigure[Loss]{
        \includegraphics[width=.4\columnwidth]{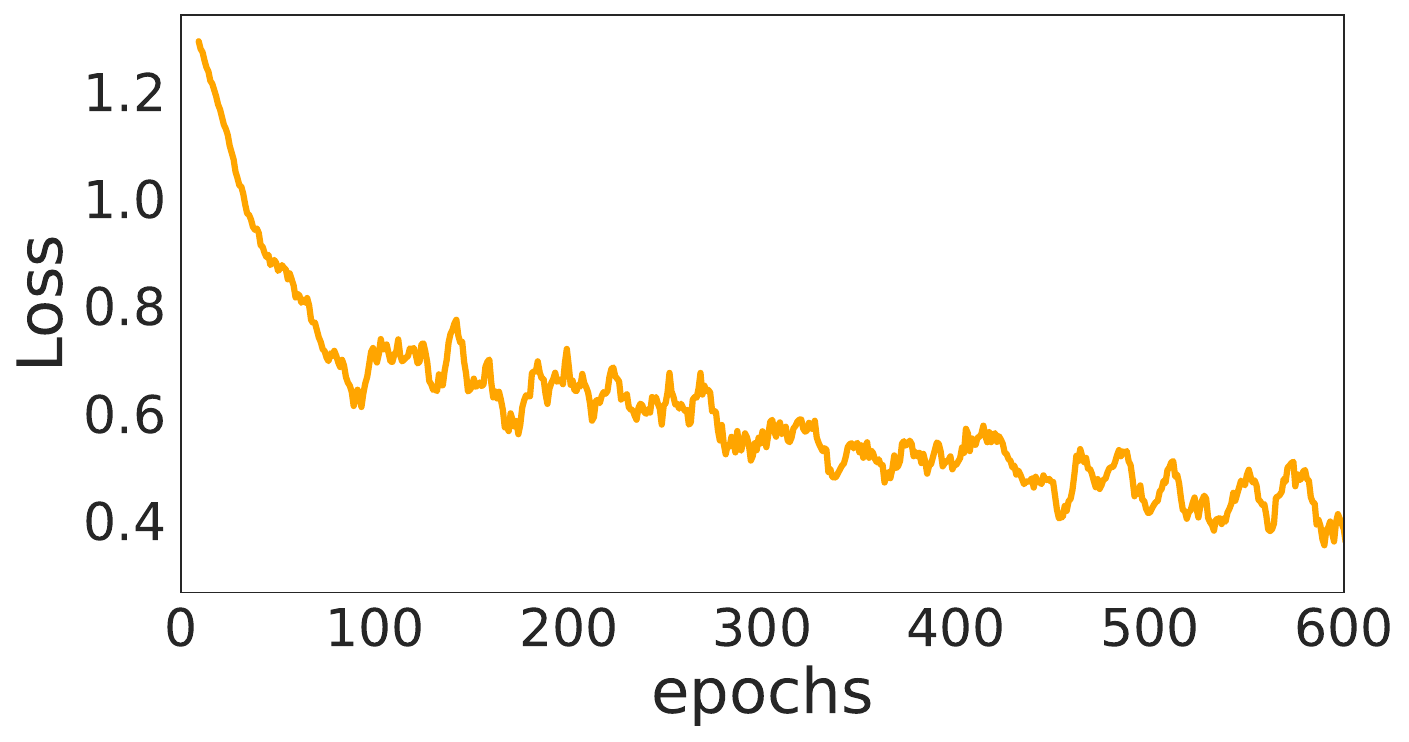}
        \label{fig:loss}
    }
    \subfigure[Reward]{
        \includegraphics[width=.4\columnwidth]{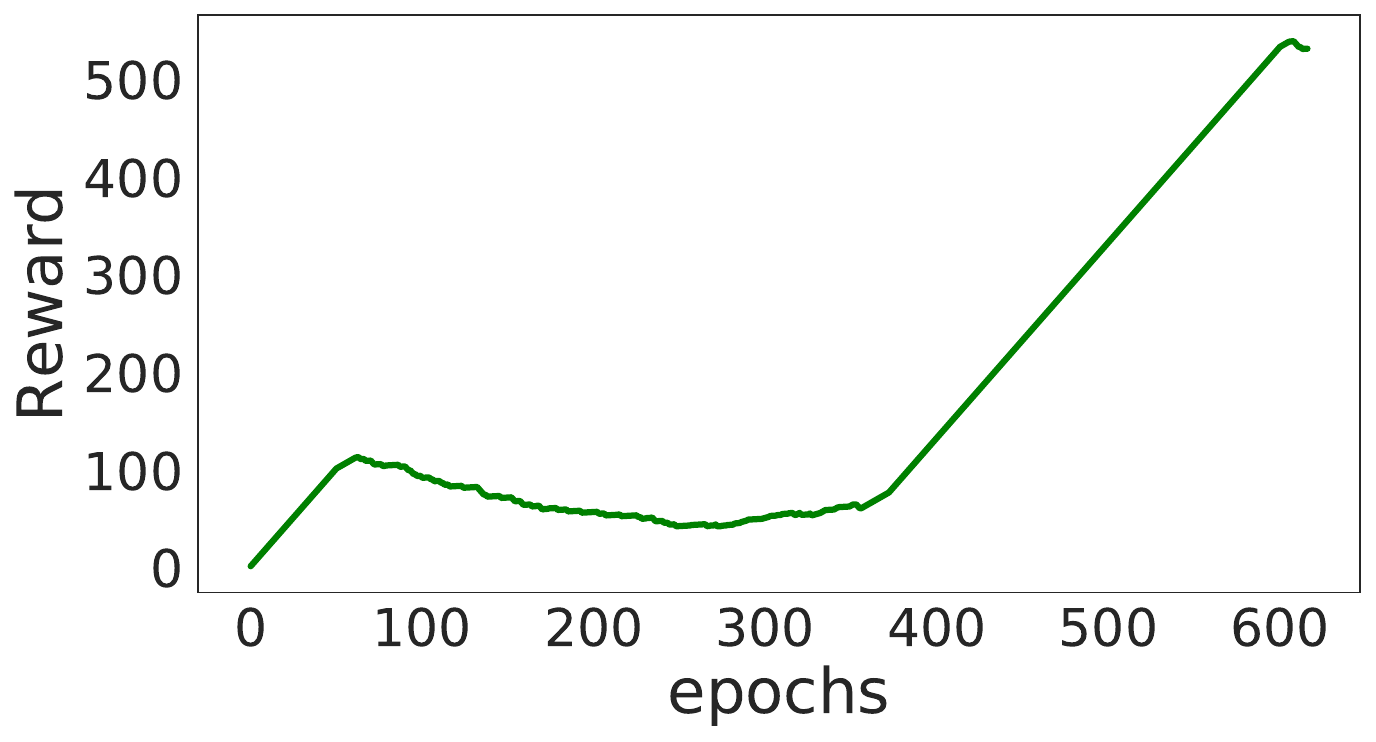} 
        \label{fig:rewards}
    }
    %\subfigure[Q-values]{
    %    \includegraphics[width=.3\columnwidth]{fig/qvalues.pdf} 
    %}
    \caption{Model performance results - Decreasing Loss and increasing Reward indicate model convergence.}
    \label{fig:DRLMeasurements}
\end{figure}

Having observed how these actions mirror the agent's interactions with the environment, we can now delve deeper into the model's performance. Figure \ref{fig:DRLMeasurements} provides an overview of the model's behavior, illustrated by the curves representing key performance metrics such as Loss and Reward.

Figure \ref{fig:loss} illustrates the trajectory of Loss throughout the experiment. A decline in Loss signifies a lower MSE in predicting q-values. In essence, a low Loss value suggests that the model is effectively learning the policy by selecting actions that maximize rewards (QoS). During the initial phase of filling the experience replay buffer, Loss tends to be higher as actions are taken without the benefit of learning, effectively representing random actions. However, as the experience replay buffer becomes populated and the Q-network is updated based on these experiences, the agent begins to make more informed and assertive decisions. This shift towards lower Loss values reflects the agent's ability to learn and improve its policy.

Turning our attention to Rewards (Figure \ref{fig:rewards}), we observe that the model incurs some penalties during the initial phase of the experiment. This corresponds to the period when the agent transitions from an initial stationary state with no charges to reaching the peak of the sinusoidal load curve, marked by the presence of 40 instances of the video player simultaneously. Subsequently, as the agent refines its decision-making, it starts receiving rewards consistently. These rewards indicate that the agent effectively maximizes the QoS of MPEG-DASH, further underscoring the model's learning and adaptive capabilities.

\begin{figure}[ht]
    \centering
    \subfigure[FPS]{
        \includegraphics[width=.45\columnwidth]{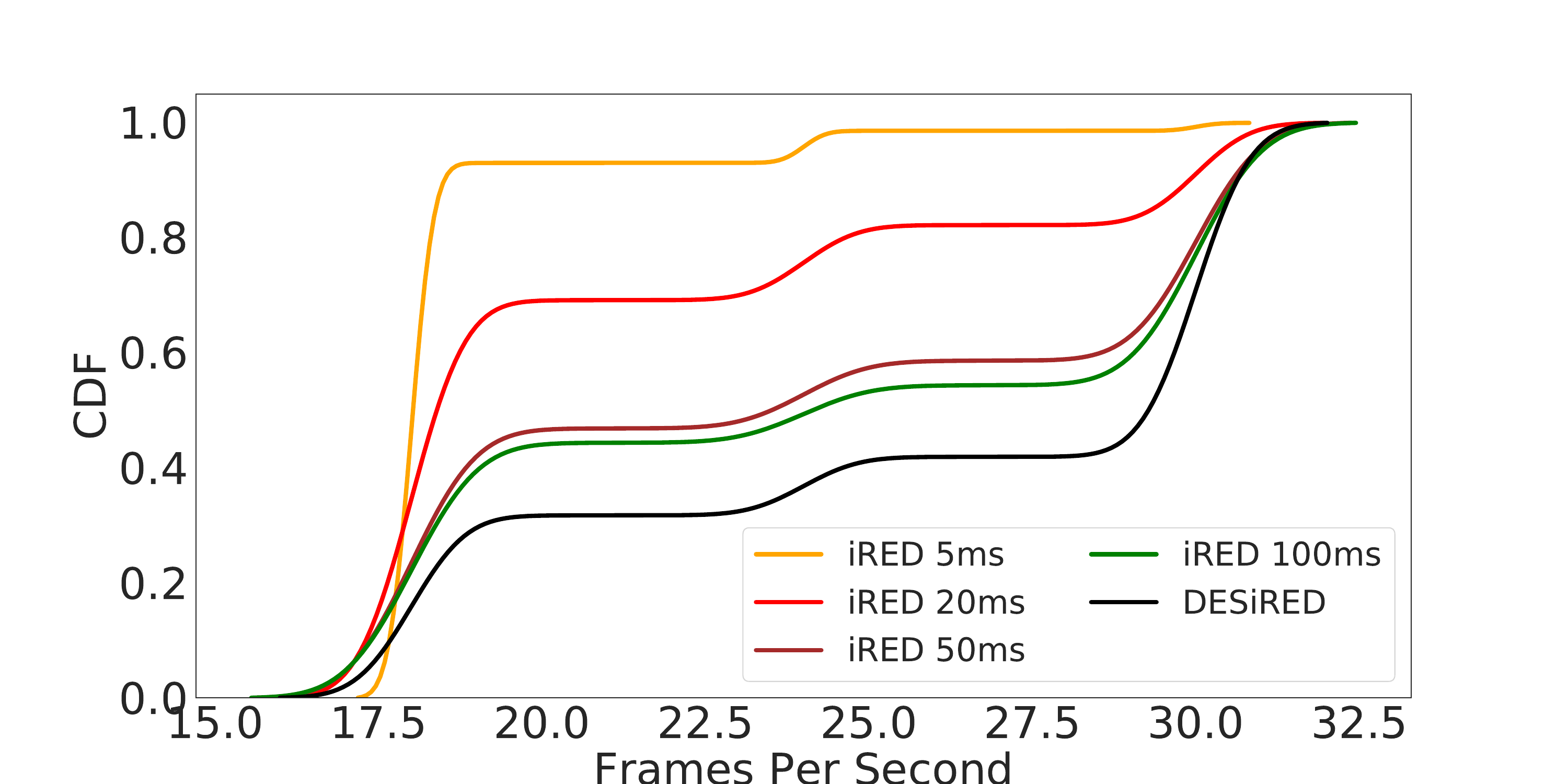}
        \label{fig:CDF_FPS_Sinusoid}
    }
    \subfigure[LBO]{
        \includegraphics[width=.45\columnwidth]{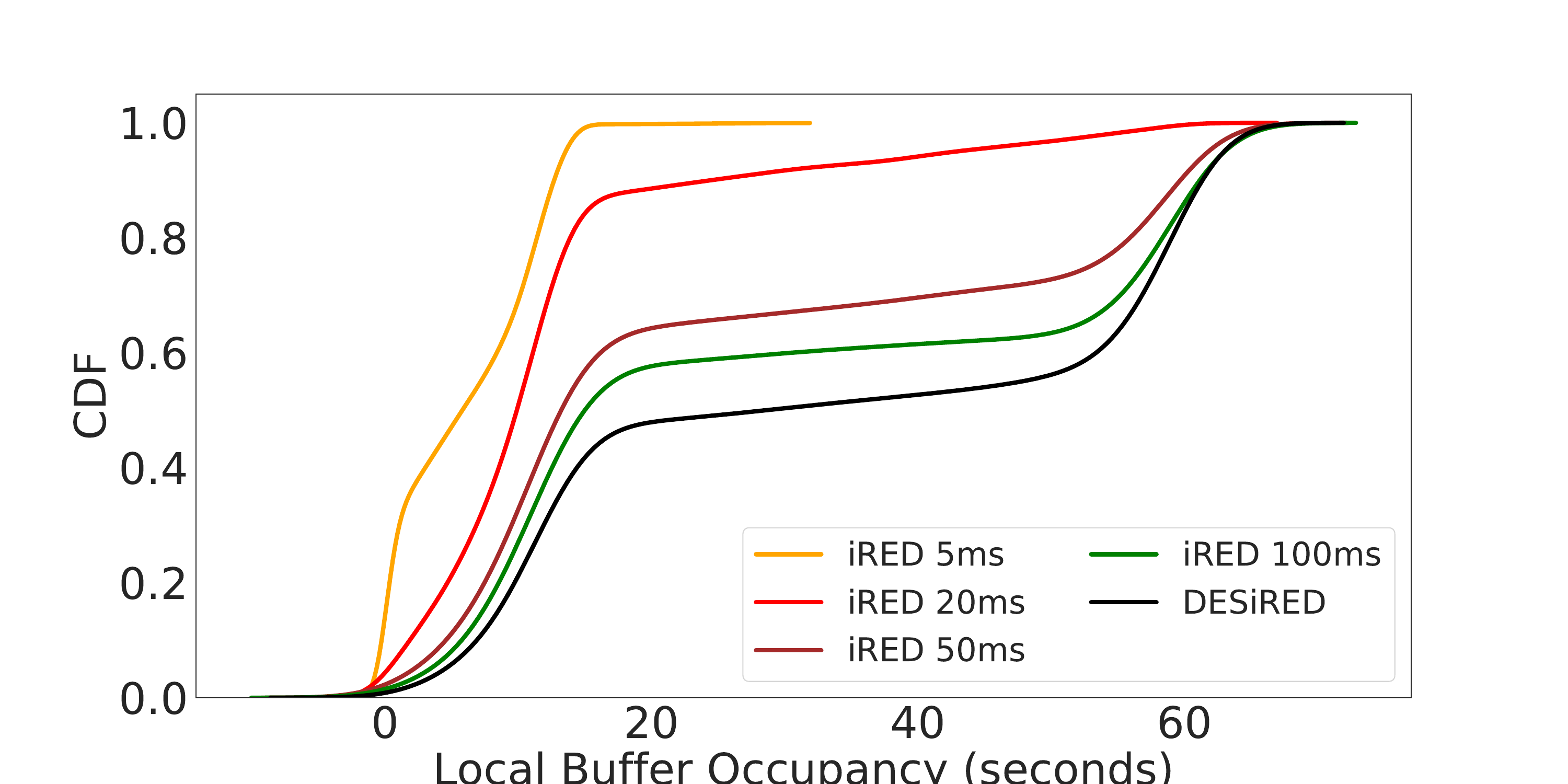}
        \label{fig:CDF_LBO_Sinusoid}
    }
    \subfigure[Video Stall]{
        \includegraphics[width=.45\columnwidth]{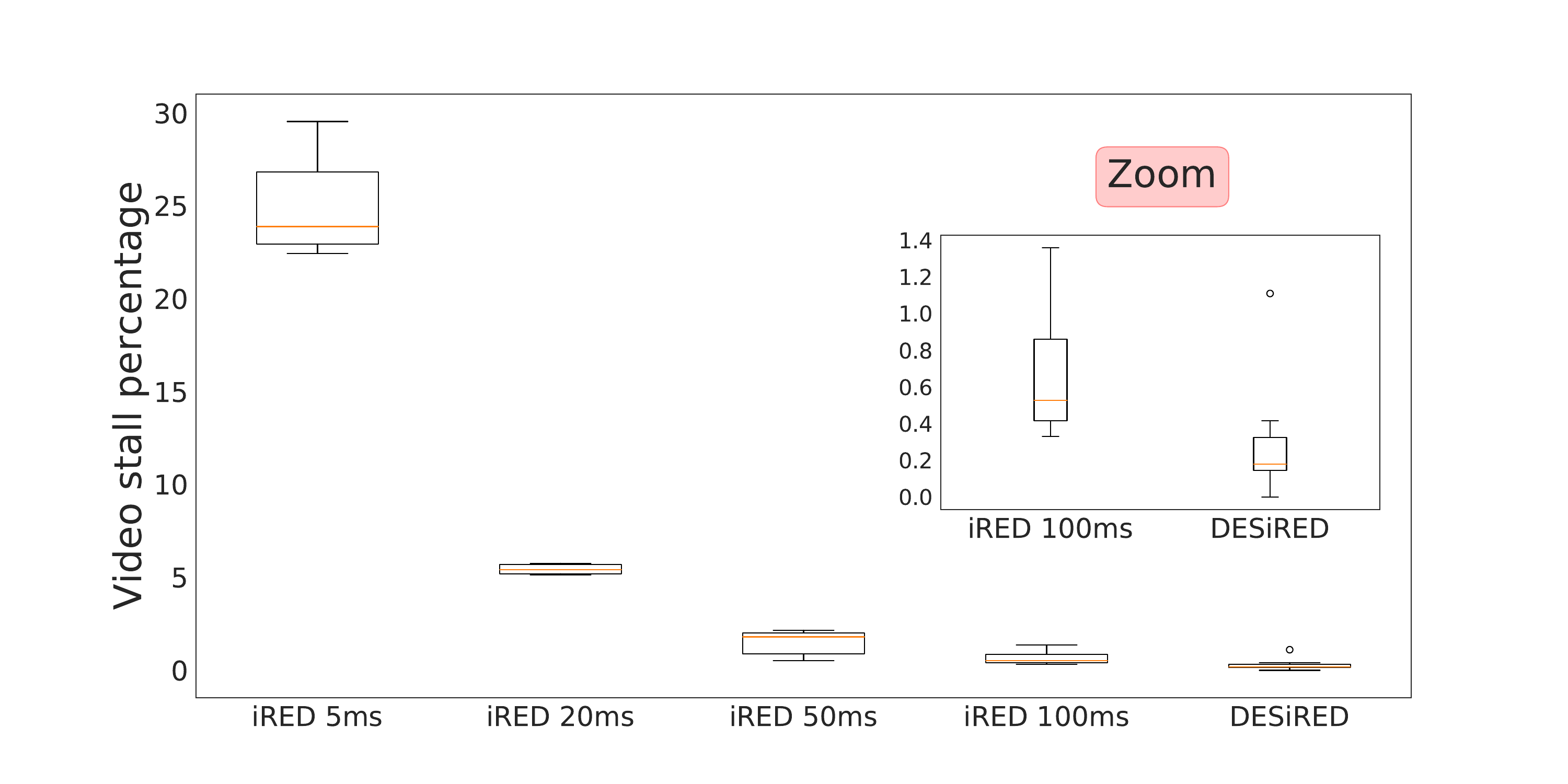}
        \label{fig:videoStall}
    }
    \caption{QoS measurements of the MPEG-DASH video service. DESiRED improves FPS and LBO while minimizing video stall.}
    \label{fig:videomeasurementsSinusoid}
\end{figure}

The insights gleaned from the agent's performance analysis are supported by the LBO and FPS metrics observed by the video client in response to DESiRED's actions, as outlined in Table \ref{tab:videoQuality} and depicted in Figure \ref{fig:videomeasurementsSinusoid}. At this conjuncture, we aim to provide an interpretation of the results from the video client's perspective, highlighting how DESiRED outperformed other approaches considered in this study.

An essential piece of data when evaluating the QoS of a video service is the resolution displayed on the screen by the video player. In this context, video consumers were offered three distinct quality levels:

\begin{enumerate}
    \item Minimum Resolution: 426x240 pixels at 18 FPS.
    \item Medium Resolution: 854x480 pixels at 24 FPS.
    \item Maximum Resolution: 1280x720 pixels at 30 FPS.
\end{enumerate}

\begin{table}[!htpb]
\centering
\scriptsize
\begin{tabular}{lcccccccc}
\toprule
\textbf{AQM} & \textbf{Min. Resolution} & \textbf{Med. Resolution} & \textbf{Max. Resolution} \\
\midrule
\textbf{iRED 5ms} & \cellcolor[gray]{0.5000}91.83\% & \cellcolor[gray]{0.9295}5.49\% & \cellcolor[gray]{0.9500}1.36\% \\
\textbf{iRED 20ms} & \cellcolor[gray]{0.6147}68.77\% & \cellcolor[gray]{0.8923}12.96\% & \cellcolor[gray]{0.8688}17.69\% \\
\textbf{iRED 50ms} & \cellcolor[gray]{0.7243}46.74\% & \cellcolor[gray]{0.8984}11.73\% & \cellcolor[gray]{0.7515}41.27\% \\
\textbf{iRED 100ms} & \cellcolor[gray]{0.7389}43.81\% & \cellcolor[gray]{0.9075}9.91\% & \cellcolor[gray]{0.7271}46.18\% \\
\textbf{DESiRED} & \cellcolor[gray]{0.7990}31.71\% & \cellcolor[gray]{0.9063}10.15\% & \cellcolor[gray]{0.6679}58.07\% \\
\bottomrule
\end{tabular}
\caption{Execution percentage at each video quality level.}
\label{tab:videoQuality}
\end{table}

Even under challenging conditions, Table \ref{tab:videoQuality} clearly demonstrates that DESiRED exhibits the highest percentage of video playback at the maximum resolution (58.07\%) and the lowest rate of playback at the minimum resolution (31.71\%). This finding aligns with the data presented in Figure \ref{fig:videomeasurementsSinusoid}.

The discussion initiated in Subsection \ref{sec:stationaryLoad} remains pertinent in this context as well. To reiterate, during periods of intense competition for shared resources, probabilistic drops facilitated by a target delay that adjusts in response to network load fluctuations have proven instrumental in maximizing the QoS of the video service. Once again, DESiRED effectively maintains a higher level of LBO filling, as depicted in Figure \ref{fig:CDF_LBO_Sinusoid}, ultimately contributing to superior FPS performance, as evidenced in Figure \ref{fig:CDF_FPS_Sinusoid}.

Figure \ref{fig:videoStall} presents a boxplot representing the percentage of video stalls, which signifies moments when the video remains frozen without any frames being displayed. A cursory glance at this figure might lead to the incorrect assumption that a longer delay at a fixed target would yield better results. However, it's important to note that DESiRED imposes an upper limit of 70ms, which is lower than the value employed by iRED100ms, thereby dispelling this theory. In this context, we believe that DESiRED's fine-tuned approach enables it to determine the optimal target delay value for each network state during the sinusoidal load.

\section{Lessons Learned} \label{sec:lessons}

In this section, we will provide insights and lessons learned from our research on applying RL to computer network problems. These insights may be valuable to the scientific community interested in using RL for similar applications.

\textit{\textbf{1) The network has an intrinsic dynamism in its behavior:}} In the realm of RL, the challenges posed by computer networks present an intriguing and multifaceted problem. In essence, an RL problem can be likened to a strategic game where an agent interacts with an environment, making decisions and receiving rewards, all within the framework of a MDP. In each of these interactions, often referred to as episodes, the agent engages in a continuous process of trial and error, striving to acquire a policy that maximizes its cumulative rewards.

However, the application of RL models to computer network-related predicaments introduces a unique set of challenges. Contemporary networks, characterized by their dynamic nature and intricate traffic dynamics, necessitate a novel approach to the integration of RL. One of the central predicaments lies in adapting an RL agent to an environment that is in perpetual flux, a paradigm well-embodied by the ever-changing states of queues within network routers.

Of notable significance is the realization that RL agents draw their learning from the experiences accumulated through their interactions with the environment. This very dependence on real-time experiences, further compounded by the interdependence between video player metrics and network conditions—themselves subject to the agent's actions—renders the use of static datasets for agent training impractical. In situations where a physical network infrastructure is not readily available, a promising alternative entails the utilization of a model capable of simulating authentic network behaviors, such as a Generative Adversarial Network (GAN) as proposed by Navidan et al. \cite{navidan2021generative}.

An additional layer of complexity is introduced through the modulation of network load patterns, a deliberate endeavor aimed at inducing the RL agent to adapt dynamically to both peak (high load) and trough (low load) network scenarios. In this pursuit, an array of network load settings was meticulously explored, encompassing flashcrowd and sinusoid patterns. Notably, the most compelling outcomes were achieved when employing sinusoidal patterns, characterized by single instances of peak and trough conditions within the duration of video streaming.

Furthermore, the intricate calibration of parameters pertaining to the reward policy emerged as an arena of paramount importance within the implementation of the DESiRED system. It was during this phase that some of the most noteworthy findings and developments transpired. Remarkably, the strategic revision of the reward policy wielded a disproportionate influence over the observed outcomes, eclipsing the impact of various other elements intrinsic to the proposed approach. As such, it underscores the pivotal significance of meticulous and judicious reward policy design tailored to the specific problem domain.

In summation, the application of RL methodologies to the domain of computer networks is an enthralling endeavor replete with challenges and opportunities. It necessitates an astute orchestration of dynamic simulations, judicious load modulation, and the nuanced refinement of reward policies—a multifaceted tapestry of considerations aimed at navigating the intricate terrain of modern network optimization.

%In this work, we experimented with several parameter adaptations, from the load pattern to the reward policy. The latter being the one that presented the most exciting results. For this reason, we believe that a reward policy must be very well planned for each problem scope.

\textit{\textbf{2) The core of solution design lies in the rewards policy:}} As previously mentioned, a significant portion of our modeling effort was dedicated to defining a rewards policy that aligns with our goal of maximizing the QoS in MPEG-DASH. Initially, we considered focusing solely on the FPS values during video playback. However, this approach proved insufficient due to the dynamic nature of the video player's adaptation logic, which considers factors like throughput and buffer level. As the agent's actions influence the target delay on network devices, we anticipated that FPS values would only exhibit noticeable changes following alterations in the LBO, as LBO is more responsive to network variations. Consequently, we opted to construct our rewards policy, with a primary emphasis on evaluating LBO levels, and secondary consideration given to FPS.

\textit{\textbf{3) Why actuate in all devices at the same time:}} As actions taken by our agent are intrinsically intertwined with the rewards policy, this study delves into several approaches, including the independent execution of actions on individual switches (each switch having its specific action) or the simultaneous execution of identical actions across all switches within the network. Initially, we contemplated that employing independent actions for each switch could be an appealing strategy. However, this approach did not align seamlessly with the scope of our problem.

Firstly, modifying the target delay for a single switch might not suffice to effectively assist the TCP congestion control algorithm, potentially yielding inconspicuous improvements in application-level QoS. Secondly, the adoption of such an approach would entail a proliferation of actions, scaling exponentially with the number of switches in the network (i.e., $2^n$ actions, with $n$ representing the count of switches). This increase in action space complexity could substantially augment the neural network architecture's intricacy.

\textit{\textbf{4) When we need to think in Transfer Learning (TL):}} Given the vast diversity of services, applications, topological configurations, and network loads encountered, it is imperative to acknowledge that an agent trained within a specific network environment cannot be expected to replicate its performance in other heterogeneous settings. In response to this challenge, TL has emerged as a promising approach, aiming to address several intricacies not typically encountered in the realm of RL. However, the application of TL within an RL framework is a non-trivial undertaking, necessitating numerous adaptations to enable the agent to effectively leverage knowledge acquired in a source domain for application in a target domain.

Amidst the inherent complexities of this context, numerous questions naturally arise, including but not limited to: a) What types of knowledge are amenable to successful transfer? b) Which RL structures are best suited for integration into a TL framework? c) What truly distinguishes a source domain from a target domain? These inquiries, among many others, prompt a comprehensive in exploration. While extant literature, such as previous work by \cite{TL_RL:2020}, has endeavored to shed light on these considerations, we posit that a dedicated examination of these issues within the specific context of Transfer Learning in RL, particularly within computer network problem domains, is required.

\section{Conclusions and Future Directions}

In summary, this study introduces DESiRED (Dynamic, Enhanced, and Smart iRED) as an innovative solution to tackle the long-standing issue of fixed target delay in AQM systems. By harnessing advanced network telemetry within programmable data planes and leveraging the capabilities of deep reinforcement learning, DESiRED emerges as a formidable tool to augment TCP congestion control mechanisms. In this novel framework, DESiRED utilizes high-resolution router buffer measurements, collected at line rate within the data plane, as inputs to deep reinforcement learning models residing on the control plane. Empowered by these synergistic components, the agent undertakes dynamic adjustments to the AQM's target delay in real-time, with the overarching goal of optimizing QoS for networked applications.

The comprehensive evaluation conducted within a realistic testbed, featuring the contemporary adaptive bitrate schemes for HTTP-based streaming (MPEG-DASH), reaffirms the viability of DESiRED. Throughout a diverse range of scenarios, encompassing various real-world traffic loads, our results consistently indicate the efficacy of dynamic target delay adjustments in enhancing the QoS of DASH video services for end users.

Considering the inherent dynamism of computer network environments, the prospect of transitioning toward TL has surfaced as a compelling avenue for future exploration. Nevertheless, the intricate challenges associated with this paradigm necessitate dedicated research endeavors to delve into these complexities in greater depth. As such, we recommend that this critical topic be addressed in forthcoming investigations.
\bibliographystyle{ieeetr}
\bibliography{main}

\end{document}